\documentclass[%
 reprint,
superscriptaddress,
 amsmath,amssymb,
 aps
]{revtex4-1}

\usepackage{graphicx}
\usepackage{dcolumn}
\usepackage{bm}
\usepackage[dvipsnames]{xcolor}
\usepackage{float}

\begin{document}

\preprint{APS/123-QED}

\title{A Kerr-microresonator optical clockwork}

\author{Tara E. Drake}\email{tara.drake@nist.gov}
\affiliation{Time and Frequency Division, National Institute of Standards and Technology, 385 Broadway, Boulder, CO 80305, USA}

\author{Travis C. Briles}
\affiliation{Time and Frequency Division, National Institute of Standards and Technology, 385 Broadway, Boulder, CO 80305, USA}
\affiliation{Department of Physics, University of Colorado, Boulder, Colorado, 80309, USA}

\author{Daryl T. Spencer}%
\affiliation{Time and Frequency Division, National Institute of Standards and Technology, 385 Broadway, Boulder, CO 80305, USA}

\author{Jordan R. Stone}
\affiliation{Time and Frequency Division, National Institute of Standards and Technology, 385 Broadway, Boulder, CO 80305, USA}
\affiliation{Department of Physics, University of Colorado, Boulder, Colorado, 80309, USA}

\author{David R. Carlson}
\author{Daniel D. Hickstein}
\affiliation{Time and Frequency Division, National Institute of Standards and Technology, 385 Broadway, Boulder, CO 80305, USA}

\author{Qing Li}
\author{Daron Westly} 
\author{Kartik Srinivasan}
\affiliation{Center for Nanoscale Science and Technology, National Institute of Standards and
Technology, Gaithersburg, Maryland 20899, USA}

\author{Scott A. Diddams}
\author{Scott B. Papp}\email{scott.papp@nist.gov}
 \affiliation{Time and Frequency Division, National Institute of Standards and Technology, 385 Broadway, Boulder, CO 80305, USA}
 \affiliation{Department of Physics, University of Colorado, Boulder, Colorado, 80309, USA}

\date{\today}

\begin{abstract}
Kerr microresonators generate interesting and useful fundamental states of electromagnetic radiation through nonlinear interactions of continuous-wave (CW) laser light. Using photonic-integration techniques, functional devices with low noise, small size, low-power consumption, scalable fabrication, and heterogeneous combinations of photonics and electronics can be realized. Kerr solitons, which stably circulate in a Kerr microresonator, have emerged as a source of coherent, ultrafast pulse trains and ultra-broadband optical-frequency combs. Using the f-2f technique, Kerr combs support carrier-envelope-offset phase stabilization for optical synthesis and metrology. In this paper, we introduce a Kerr-microresonator optical clockwork based on optical-frequency division (OFD), which is a powerful technique to transfer the fractional-frequency stability of an optical clock to a lower frequency electronic clock signal. The clockwork presented here is based on a silicon-nitride (Si$_3$N$_4$) microresonator that supports an optical-frequency comb composed of soliton pulses at 1 THz repetition rate. By electro-optic phase modulation of the entire Si$_3$N$_4$ comb, we arbitrarily generate additional CW modes between the Si$_3$N$_4$ comb modes; operationally, this reduces the pulse train repetition frequency and can be used to implement OFD to the microwave domain. Our experiments characterize the residual frequency noise of this Kerr-microresonator clockwork to one part in $10^{17}$, which opens the possibility of using Kerr combs with high performance optical clocks. In addition, the photonic integration and 1 THz resolution of the Si$_3$N$_4$ frequency comb makes it appealing for broadband, low-resolution liquid-phase absorption spectroscopy, which we demonstrate with near infrared measurements of water, lipids, and organic solvents.


\end{abstract}

\pacs{Does PRX still use PACS numbers?}
\maketitle


\section{Introduction}
Optical-atomic clocks 
\cite{Ludlow2015}, which are among the most precise metrological instruments currently available, provide continuous-wave (CW) laser radiation that is stabilized to a narrow-linewidth atomic transition. Optical-frequency combs provide a clockwork to coherently transfer the stability of an optical clock to all the comb modes and to a microwave signal derived from the comb repetition frequency. The latter capability, called optical-frequency division (OFD), was developed for optical timekeeping \cite{Diddams2001,Diddams2000_1}, and it leverages the frequency multiplication inherent in the comb's mode spectrum, namely $\nu_n = f_\textrm{ceo} + n\,f_\textrm{rep}$ where $f_\textrm{ceo}$ is the carrier-envelope offset frequency and $f_\textrm{rep}$ is the repetition frequency. To implement OFD, mode $n$ of an optical frequency comb is phase-locked to an optical reference such that $f_{\textrm{rep}} = (\nu_n-f_\textrm{ceo})/n$, hence dividing the optical clock carrier frequency to a lower frequency by a factor of $n$ and concurrently reducing the phase noise by approximately the factor $20\,\log_{10}(n)$. High-performance tabletop oscillators make use of this property \cite{Bartels2005,McFerran2005,Millo2009,Weyers2009,Zhang2010,Fortier2011,Baynes2015,Xie2016}.

\begin{figure*}[ht]
	\centering
	 \includegraphics[width=0.7\linewidth]{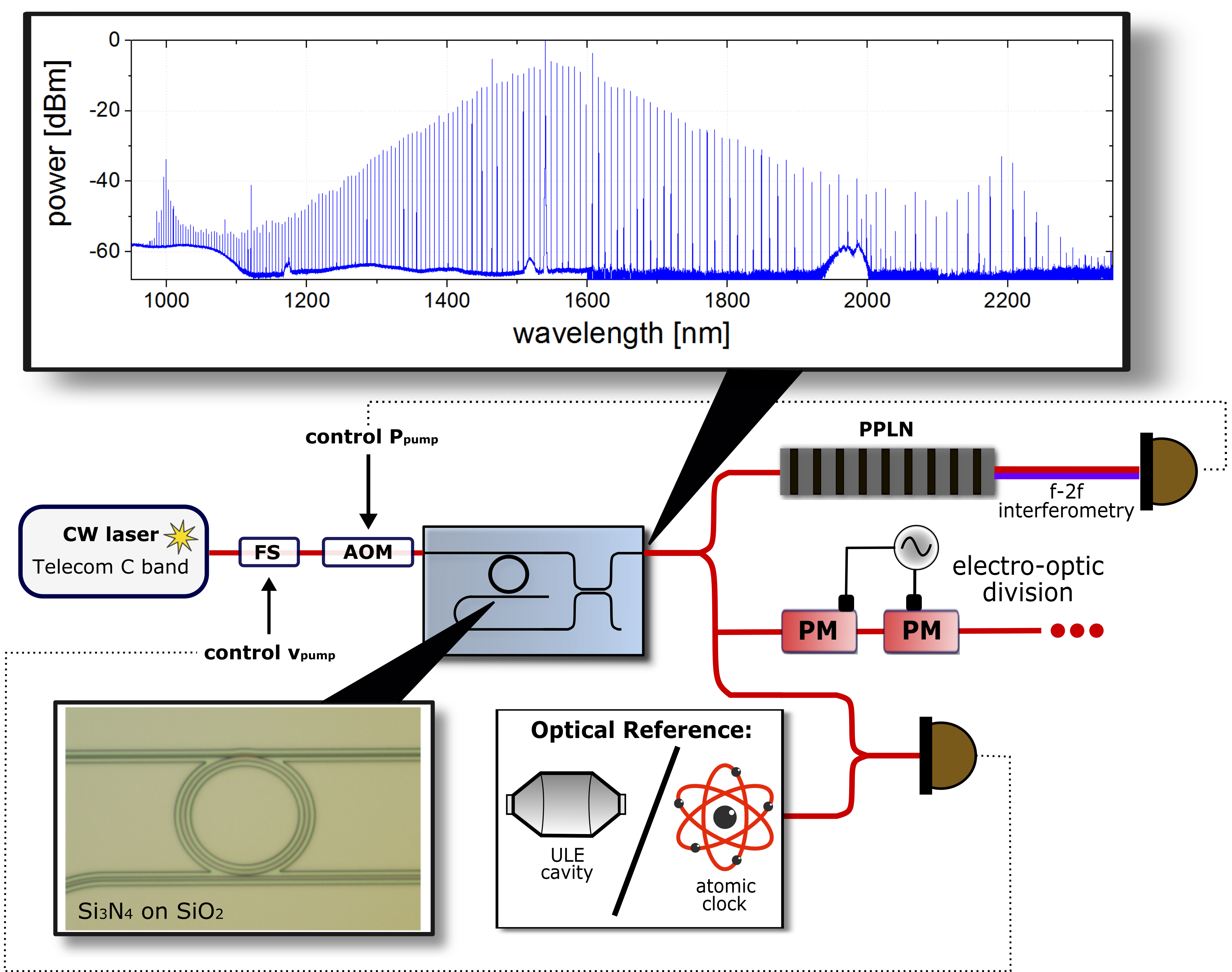}
\caption{Schematic of the Kerr-microresonator clockwork. The microresonator is a Si$_3$N$_4$ ring with radius of 23 $\mu$m. It is pumped with a CW laser that is coupled into a 720 nm wide access waveguide (upper waveguide in this picture) via lensed fibers. The resultant comb spans more than an octave, and wavelengths from 900--1900 nm are outcoupled from the resonator via the access waveguide through port while wavelengths $>$1900 nm are outcoupled via a drop port to a 1200 nm wide waveguide (lower waveguide in this picture). The comb light is recombined via an adiabatic dichroic coupler before leaving the chip. The microcomb is stabilized through f-2f interferometry and locking to an optical-frequency reference. The f-2f technique is carried out with a waveguide PPLN and a PID feedback circuit controlling the CW pump intensity via an acousto-optic modulator (AOM). The optical reference (a ULE-cavity-stabilized laser in this paper) creates a heterodyne beat with one of the comb modes, and a second PID feedback circuit controls the CW pump frequency via a QPSK modulator (FS). When both PID feedback circuits are operational, the fractional stability of the optical cavity is transferred to $f_\textrm{rep}$, using OFD. The full comb spectrum is also sent through a series of electro-optic phase modulators (PM) to reduce the effective repetition frequency.} \label{fig:device}
\end{figure*}

Dissipative-soliton generation in Kerr-microresonators \cite{Kippenbergeaan8083} is a recent advance that enables the implementation of modelocked frequency combs in miniature devices, with relatively low power consumption and the potential for planar integration. This work builds on decades of knowledge in soliton nonlinear optics \cite{AgrawalBook}. The development of Kerr-soliton microresonator frequency combs (``microcombs'' in this paper) has led to explorations of novel nonlinear states of light \cite{Cole2017,Yao2018}, microresonator soliton \cite{Li2016,Yi2015}, and demonstrations of functional devices such as clocks \cite{Papp2014}, optical synthesizers \cite{Briles2017,Spencer2017}, dual-comb spectrometers \cite{Dutt2016,Suh2016}, massively parallel communications \cite{Marin-Palomo2017}, and high-speed ranging \cite{Trocha2017}. Recent experiments have developed access to the carrier-envelope offset frequency of microcombs, $f_\textrm{ceo} = \nu_p - N\,f_\textrm{rep}$, where $\nu_p$ is the pump laser frequency and $N$ is the comb mode number to measure $f_\textrm{ceo}<f_\textrm{rep}$, through 2f-3f \cite{Jost2014,Brasch2017} and f-2f measurements \cite{Briles2017,Spencer2017}. 

Developing integrated-photonics frequency combs for OFD will enable new applications of optical timekeeping. The recently developed very high $f_\textrm{rep}\approx 0.2-1$ THz soliton combs \cite{Li2016,Pfeiffer2017} that leverage careful engineering of the waveguide dimensions for spectral bandwidth and $f_\textrm{ceo}$ control \cite{Briles2017} are useful for microwave, mm-wave, and THz photonic-microwave generation \cite{Endo2018}. Phase stabilization of $f_\textrm{ceo}$ is unavoidably required, since real time tracking and correction for the comb repetition frequency-- the output of the clockwork whose stability ought to exceed existing microwave oscillators --is not possible. 
Here we report a Kerr-microresonator optical clockwork that implements OFD to accurately generate a microwave-frequency output directly from an optical clock. Our experiments use an f-2f self-referenced Si$_3$N$_4$ microcomb and $f_\textrm{ceo}$ phase stabilization. With these tools, we show how to perfectly divide the frequency and transfer the stability of an optical-clock laser to $f_\textrm{rep}$ of the Si$_3$N$_4$ microcomb. We verify the fractional-frequency accuracy and precision of this Kerr-microresonator clockwork to the 10$^{-17}$ level after continuous, glitch-free operation for two hours. This experiment is enabled by measurements with respect to an OFD system based on a f-2f self-referenced electro-optic (EO) modulation comb (hereafter referred to as the "reference comb" for clarity) \cite{Carlson2017}. Comparison of these two OFD clockworks jointly evaluates their residual stability, including the optical and microwave network that interfaces them. A further benefit of combining microcomb and EOM-comb technology is that direct EO modulation of our Si$_3$N$_4$ comb effectively reduces its mode spacing to an electronically detectable rational sub-multiple of 1 THz. With this technique, we demonstrate 33 GHz or 8 GHz mode spacing combs with up to 43 THz of optical bandwidth. An additional experiment with our octave-spanning Kerr comb demonstrates practical absorption spectroscopy in the near-infrared (NIR) wavelength range of four liquids that are important to biological, medical, laboratory, and food sciences. 

\begin{figure}[t!]\vspace{-12pt}
\centering
\includegraphics[width=1.04\linewidth]{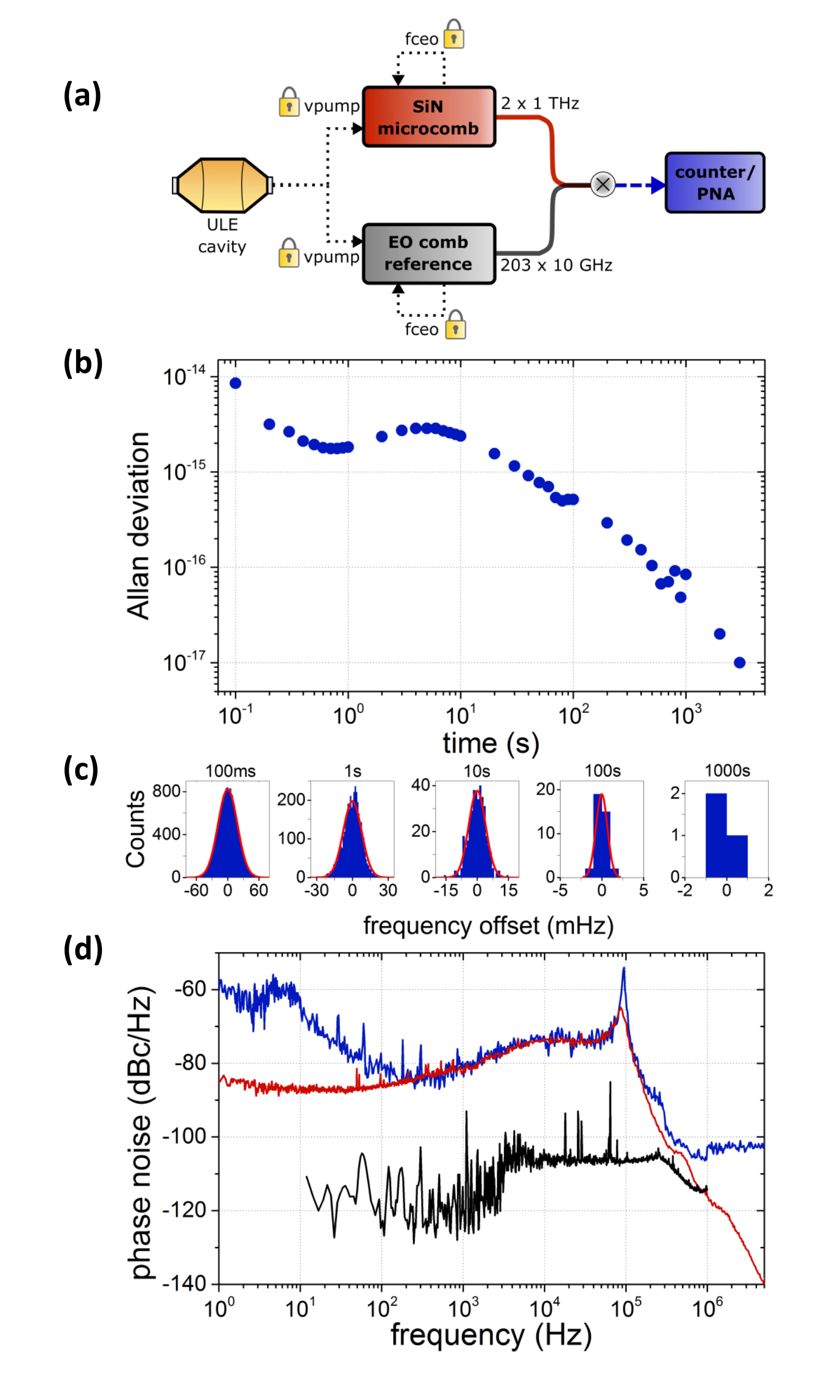}
\caption{Studying the residual noise of the Kerr-microresonator clockwork. (a) The principle of our measurements is parallel operation of two OFDs, the microcomb clockwork and the EOM reference-comb OFD. We measure and subtract optical-heterodyne RF beats between the combs at 1540 nm and 1524 nm, which characterizes this precise difference in the combs' repetition frequencies: $f_\textrm{diff} = 2*f_\textrm{rep SiN} - 203*f_\textrm{rep EO}$. (b) Allan deviation of $f_\textrm{diff}$. (c) Histograms of a two-hour measurement of $f_\textrm{diff}$ relating the absolute frequency accuracy of the two OFDs by direct comparison to its expected value. (d) Phase-noise spectrum of $f_\textrm{diff}$ (blue), $f_\textrm{ceo}$ (red), and the reference comb $f_\textrm{rep EO}$ (black), indicating that reference OFD contribution is not significant.}
\label{fig:OFD_RN} \end{figure} \vspace{-12pt}

\vspace{-0pt}\section{Kerr-microresonator clockwork}
Figure \ref{fig:device} shows our Kerr-microresonator clockwork setup. The microresonator is a silicon-nitride (Si$_3$N$_4$ hereafter truncated to SiN) microring, fabricated by way of low-pressure chemical-vapor deposition (LPCVD), electron-beam lithography, and chemical etching \cite{Li2016}. The device has an unloaded mode quality factor of $\approx$2,500,000 and a radius of $\approx$23 $\mu$m. The SiN waveguide is air clad on three sides, and its width, height, and radius are carefully chosen for the desired dispersion and carrier-envelope-offset frequency \cite{Li2016,Briles2017}. The resonator is pumped with $\approx$200 mW of chip-coupled CW laser power from an 1540 nm external-cavity diode laser (ECDL); the frequency and power of the pump laser is controlled using a frequency shifter (FS) and acousto-optic modulator (AOM), respectively. The FS enables laser frequency sweeps across a SiN ring resonance that are faster than the thermal heating rate of the device, thereby avoiding a large thermal bistability and making soliton formation rather straightforward; see \cite{Stone2017} and \cite{Briles2017} for more details. The dispersion of the SiN resonator enables generation of a soliton microcomb with modes exceeding 1 nW (often 10 nW) of optical power per mode from 960 nm to 2300 nm, and the power at either extreme is increased by dispersive waves \cite{Briles2017}; see Fig. \ref{fig:device}. 

To introduce the general principle of clockwork operation according to the equation $f_\textrm{rep} = (\nu_p - f_\textrm{ceo})/N$, we phase-lock one of the microcomb's modes to an optical-frequency reference and simultaneously stabilize $f_\textrm{ceo}$. In this paper, the optical-frequency reference is provided by a CW laser that is locked via the Pound-Drever-Hall method to a single mode of an ultra-low-expansion (ULE) cavity \cite{Baynes2015}. This cavity-stabilized laser achieves a fractional-frequency stability of $\approx10^{-15}$ for 1 second measurements, and it has a typical drift rate of 100 mHz/s. In particular, the central microcomb mode, corresponding to the pump laser, is phase-locked to the cavity-stabilized laser via an electronic feedback circuit to the FS. The choice of optical reference is flexible; another possibility is to use a compact atomic reference based on the two-photon Rb transition at 778 nm (Newman, et al., in preparation).

We phase-lock $f_\textrm{ceo}$, obtained using the f-2f technique, to a hydrogen-maser-referenced microwave oscillator, using the pump-laser intensity. The result is phase-coherent conversion of the optical reference to $f_\textrm{rep}$ of the microcomb. Since full integration of the SiN microresonator and PPLN waveguide on the same chip has not yet been achieved, the doubling process must be done after outcoupling light from the chip with the SiN resonator and re-coupling onto the chip with the PPLN waveguide. The loss associated with these processes ($\approx13$ dB total), the efficiency of the PPLN ($\approx 30\%$ per W on chip), and the scaling of the doubling process with the square of the intensity of the 1998 nm light all conspire to make direct doubling of 1998nm comb mode infeasible. Instead, we send an auxiliary laser (Newport TLB-6736 ECDL, wavelength of 1998 nm) through the PPLN waveguide to collect 10 mW of 1998 nm light and 0.03 mW 999 nm. Two optical heterodyne are formed with the comb modes at 1998 nm and 999 nm, and these two frequencies are mixed together with a 2:1 ratio to directly measure $f_\textrm{ceo}$. 

\vspace{-9pt}
\section{Characterizing the Kerr-microresonator clockwork}
With operational microcomb OFD, we characterize the residual-noise contribution of the Kerr-microresonator clockwork to its optical-frequency reference. We perform these measurements by comparing the result of OFD-- the 1 THz repetition frequency of the microcomb --to a reference comb that also implements OFD of the same ULE cavity-stabilized laser. In this frequency comparison of two 1-THz signals, the residual noise we measure indicates the contribution of both OFD systems, and the microwave and optical network that connects them. Our reference OFD system is an EOM frequency comb with 10 GHz repetition frequency that is stabilized by supercontinuum generation with a SiN photonic-chip waveguide and f-2f interferometry. Reference \cite{Carlson2017} provides a detailed functional explanation of the reference EOM-frequency-comb OFD.

Figure 2a shows the concept of the residual-noise measurement. The microcomb OFD and the reference-comb OFD operate in parallel, and to compare them, we measure the 200-th harmonic of the reference-comb repetition frequency with respect to the 2nd harmonic of the microcomb repetition frequency. We obtain this comparison of the two combs directly, without need to detect any signals at 2 THz, by forming optical-heterodyne beats of the reference comb and microcomb modes at 1540 nm and 1524 nm, respectively. The difference of these heterodyne signals, which we obtain by electronic frequency mixing, directly reflects the difference of the two OFDs; we analyze this signal with a frequency counter and a phase-noise analyzer.

In practice, the two comb systems exist in separate labs at NIST, and the reference comb is delivered to the microcomb lab through $\sim200$ meters of optical fiber. We use an optical phase-lock of the microcomb pump laser to one mode of the reference comb. This configuration allows for an almost perfect cancellation of the ULE cavity laser's frequency drift; see the supplementary material for more information on how we process our data. The result of our residual-noise measurement is a frequency-counter record and phase-noise analysis of $f_\textrm{diff} = 2*f_\textrm{rep SiN} - 203*f_\textrm{rep EO}$, which are shown in Fig. \ref{fig:OFD_RN}b-d and present a detailed picture of our clockwork's performance. Here $f_\textrm{diff}$ is a quantity that we can predict based on the frequency setpoints of our phase-locked combs. The Supplementary Materials contain more information regarding the measurement setup.

We verify that the two OFD systems are accurate by subtracting our measurements of $f_\textrm{diff}$ from its expected value; see Fig. \ref{fig:OFD_RN}c. A zero frequency mean of the histograms in Fig. \ref{fig:OFD_RN}c, within an uncertainty indicated by the Allan deviation presented below in Fig. \ref{fig:OFD_RN}b, indicates. Since the two OFDs share a common optical reference, $f_\textrm{diff}$ of their divider outputs can in principle be independent. However, since the division ratios, $N=192$ for the SiN comb and $N=19,339$ for the EOM comb, are not commensurate and cannot be straightforwardly set this way in our current experiment \cite{Carlson2017}, $f_\textrm{diff}$ drifts in time at less than 1.5 $\mu$Hz/s according to the absolute optical-frequency drift of the ULE-cavity-stabilized laser. We track this drift independently and synchronously by frequency counting $f_\textrm{rep EO}$ with respect to a hydrogen-maser reference; see the Supporting Information for more details. Without any assumption beyond the integer OFD ratios, we correct the measured $f_\textrm{diff}$. The data are presented as histograms with different timing bins in Fig. \ref{fig:OFD_RN}c, and the mean error in $f_\textrm{diff}$ from the measured $f_\textrm{rep EO}$ value (zero on the Fig. \ref{fig:OFD_RN}c x-axis) is -25 $\mu$Hz $\pm$ 95 $\mu$Hz.

Our measurements capture the residual noise of the Kerr clockwork, as described above. Specifically, Fig. \ref{fig:OFD_RN}b indicates the Allan deviation of the two parallel OFD systems with the measured cavity drift removed. From this measurement, we show that the residual noise of the microcomb clockwork is 1 part in $10^{15}$ for 1 second measurement durations and reduces to 1 part in $10^{17}$ after 2 hours of measurement (Fig. \ref{fig:OFD_RN}b), which to our knowledge is the initial demonstration of complete optical-to-microwave OFD with a microcomb and is a factor of thirty improvement over the most precise microcomb measurements reported to date \cite{Briles2017}. Presumably the Allan deviation performance plateau for 1-10 second measurements contains contributions from environmental fluctuations in the optical network that connects the two OFD systems, since we did not actively stabilize path lengths. Moreover owing to the substantial $f_\textrm{ceo}\sim12$ GHz, the hydrogen-maser-derived reference frequency for phase locking $f_\textrm{ceo}$ presumably adds noise-- this represents a fundamental challenge of our high $f_\textrm{rep}$ microcomb system design. Still, as our measurements show, microcombs can achieve high accuracy and precision.

We also present the phase-noise spectrum of $f_\textrm{diff}$ (Fig. \ref{fig:OFD_RN}d), which provides detailed insight into the performance of our clockwork. Three traces characterize this data, including the negligible contribution of the reference comb (black trace), the noise of $f_\textrm{diff}$ itself (blue trace), and the noise of $f_\textrm{ceo}$ (red trace) from which we can also infer the phase-noise spectrum of $f_\textrm{diff}$. Note that we obtain these spectra while the OFDs are phase-locked, as described above; thus they indicate a residual-phase-noise contribution within our servo bandwidth of $\approx100$ kHz. In our current configuration, the repetition-frequency noise is dominant, indicated by the agreement of the $f_\textrm{rep}$ and $f_\textrm{ceo}$ data when $f_\textrm{rep}$ is scaled by the appropriate factor of $N=192$. Therefore, the residual phase-noise contribution of the microcomb pump laser phase locking is relatively low. Understanding why $f_\textrm{rep}$ is the dominant contribution here is an important future task for microcomb experiments and applications.

\begin{figure}[ht]
	\includegraphics[width=\linewidth]{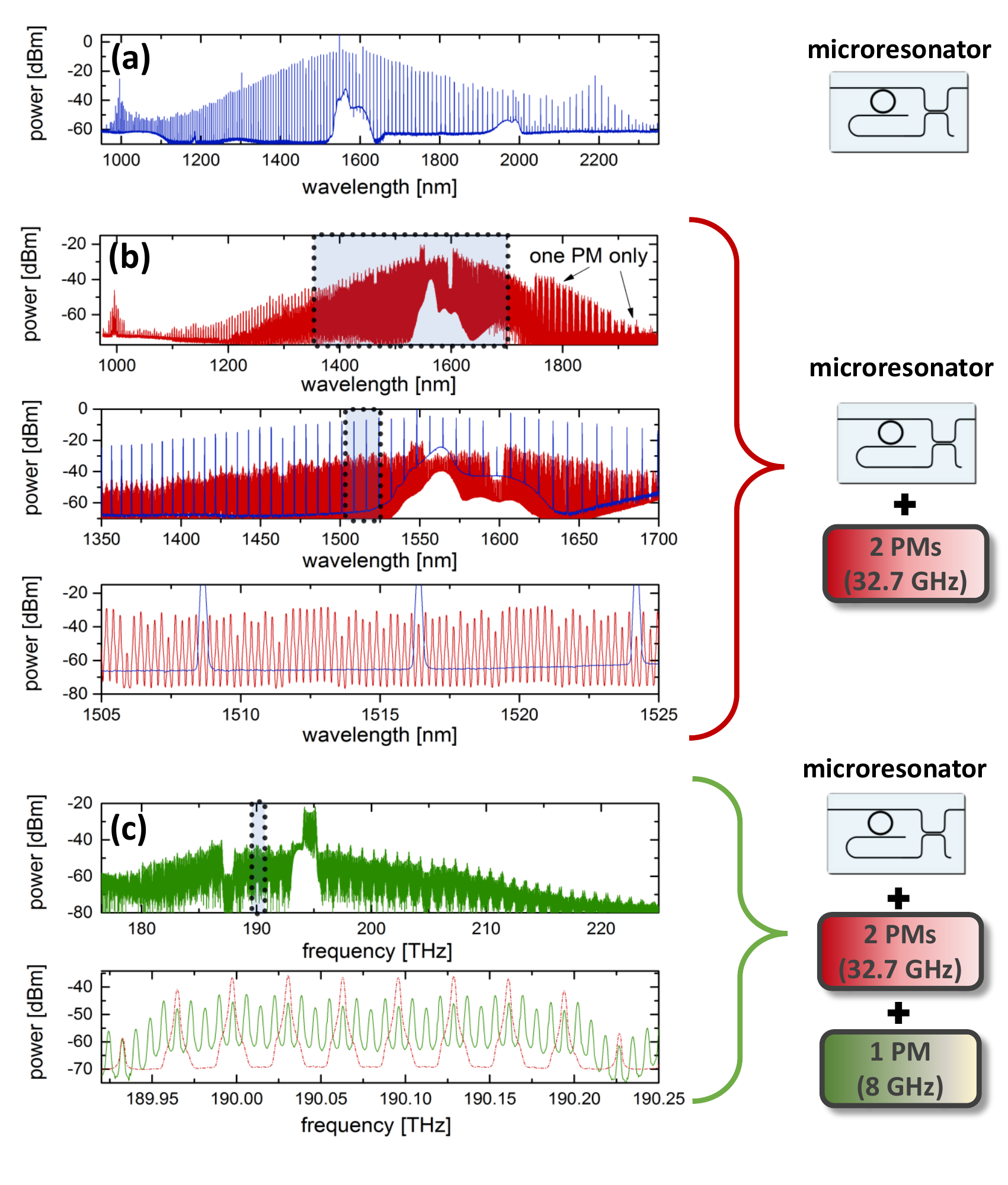}
\caption{Reduction of the 1-THz SiN microcomb mode spacing through EO phase modulation. (a) Microcomb spectrum coupled out of the SiN waveguide, using lensed fibers. (b) The full microcomb spectrum (blue) is sent through two low-$V_{\pi}$ electro-optic phase modulators (PM), which are driven at roughly 32.7 GHz ($f_\textrm{rep}$/31). These modulators are optimized for operation at 1550 nm, but they perform well over a relatively large bandwidth, and the resulting spectrum includes a filled in 32.7 GHz comb that spans nearly 350 nm (red). (c) Demonstrating further reduction, we send the 32.7 GHz comb (red) through an additional PM driven at 8 GHz. The mode spacing of the resulting spectrum (green) has been divided by 124 from the original microcomb $f_\textrm{rep}$. (This spectrum is shown in frequency units in contrast to the previous panels.)}\label{fig:EOsampling}
\end{figure}

\vspace{-9pt}\section{Microcomb repetition-rate reduction from 1 THz}
The benefits of our microcomb are an octave-spanning output spectrum, relatively high output power per mode at relatively low pump power, and photonic-chip integration. However, the 1-THz soliton pulse train is challenging to measure with a photodetector and microwave electronics. Additionally, such a widely spaced comb isn't ideal for some applications because the broad overall spectral coverage is relatively undersampled. Here we show how to address these shortcomings through the use of EO phase modulators \cite{delhaye2012}, which effectively reduce the 1-THz repetition frequency to an in principle arbitrarily low value. This is a convenient and efficient technique to build a user-defined frequency comb from our base octave-span microcomb.
 
Figure \ref{fig:EOsampling} shows the results of reducing the microcomb repetition frequency. The EO modulators generate nearly arbitrarily spaced interleaving combs about each SiN microcomb mode. Figure \ref{fig:EOsampling}a shows the full spectrum outcoupled from the microresonator and output from the photonic chip, a fraction of which is sent through two low $V_\pi$ phase modulators (figure \ref{fig:EOsampling}b). By driving these modulators at roughly $f_\textrm{rep}/31$ (in figure \ref{fig:EOsampling}b, $f_\textrm{EO}$ = 32.6749009 GHz), we generate EO subcombs that are nearly subharmonic of the microcomb $f_\textrm{rep SiN}$; the difference between $f_\textrm{rep SiN}$ and $f_\textrm{rep EO}$ yields a low frequency ($\sim$10 MHz) heterodyne beatnote between adjacent subcombs. In the case of OFD, the modulators facilitate readout of the clockwork. But this setup can also be used in a feedback loop to stabilize $f_\textrm{rep SiN}$ directly. In this alternative method of operation, the stability of the RF reference driving the modulators is transferred to the stability of each microcomb mode.

Additional modulators can be included (Fig. \ref{fig:EOsampling}c) to further subdivide the mode spacing. These provide extra tunability of the optical mode frequencies without sacrificing the microcomb $f_\textrm{rep}$ readout capability. In Figs. \ref{fig:EOsampling}b and c, the shaded regions on upper panels correspond to the enlarged spectrum in the panel below. Here we demonstrate continuous EO subcombs with a mode spacing as low as 8 GHz, which is just above the resolution of our optical spectrum analyzer, spanning up to 43 THz in the 1550 nm wavelength region. This capability, which could be straightforwardly refined for user-defined applications beyond the demonstrations reported here, is a powerful approach for experiments ranging from optical communications, dual-comb measurements, calibration of astronomical spectrographs, sensitive optical detection of signals, and various other directions. Furthermore, though we did not take the time to develop this in our experiment, intelligent referencing of the RF synthesizers to a signal derived from the $f_\textrm{ceo}$ of the microcomb could result in further OFD of the fractional stability of the ULE cavity to the 33 or 8 GHz mode spacing.

\vspace{-9pt}\section{Direct microcomb spectroscopy across the near-infrared} 
The near-infrared (NIR) region contains the first and second overtones of the fundamental vibrations of O-H, C-H, N-H, and S-H bonds and can be a complex but rich source of compositional information \cite{Wilson2015}. NIR spectroscopy enjoys extensive use in the food, agriculture, pharmaceutical, medical, chemical, and polymer industries \cite{Manley2014,Burgess2007,Rohe1999} and has been employed as a non-destructive imaging and analysis technique in archeology \cite{Tsuchikawa2005}, Martian geology \cite{Jensen2011}, and art restoration \cite{Delaney2005,Dooley2013}. Recently, attention has been given to NIR comb spectroscopy for the application of biomedical imaging, as water, lipids, and collagen, the fundamental components of biological tissue, display major absorption peaks in the 900-2300nm wavelength region \cite{Wilson2015}. In addition, microcomb devices may have the potential to perform NIR dual-comb spectroscopy \cite{Suh2016,Dutt2016}, depending on system requirements such as resolution, spectral coverage, and absolute frequency calibration.

\begin{figure}[H]
	\centering
    \includegraphics[width=\linewidth]{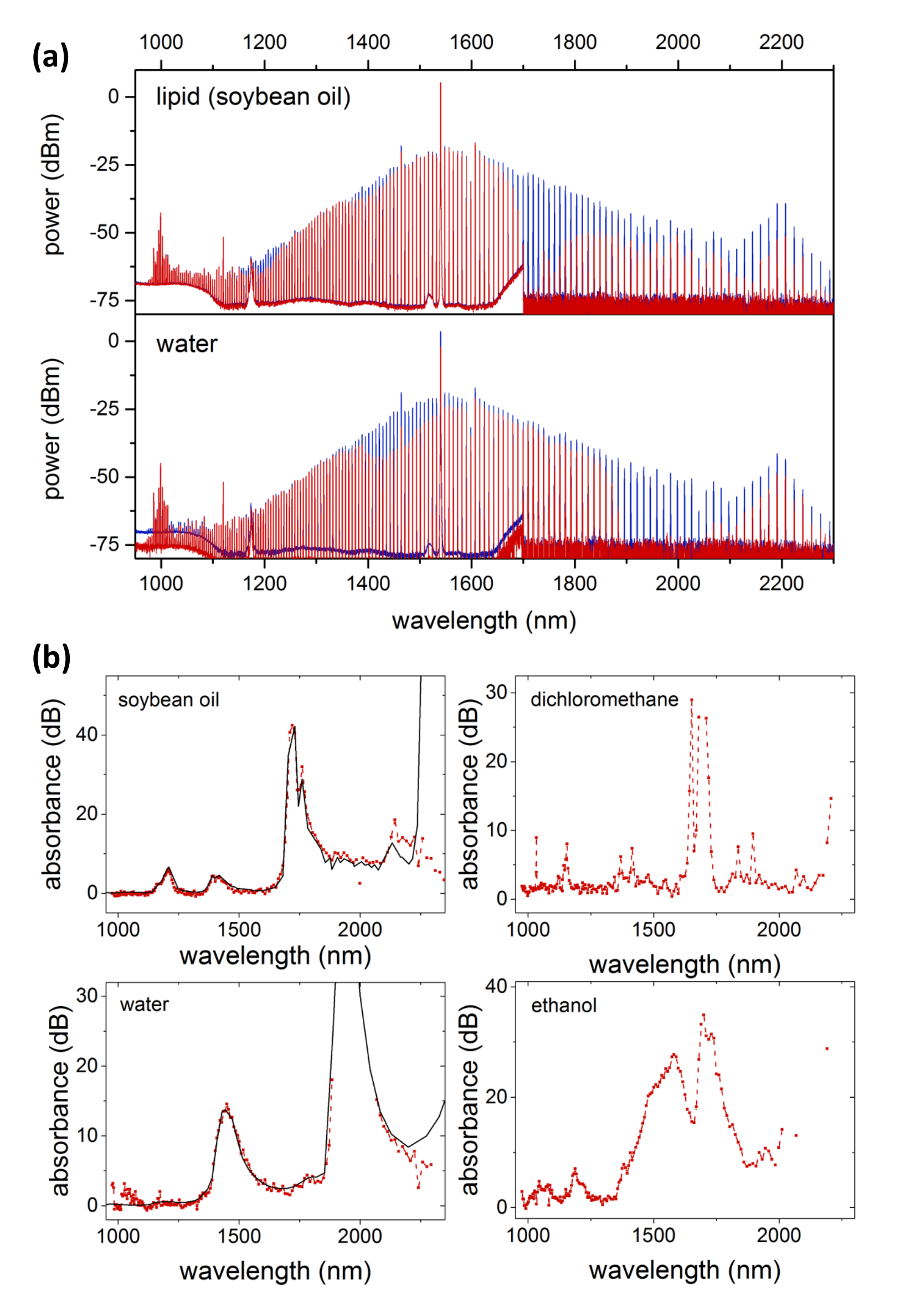}
    \includegraphics[width=1.1\linewidth]{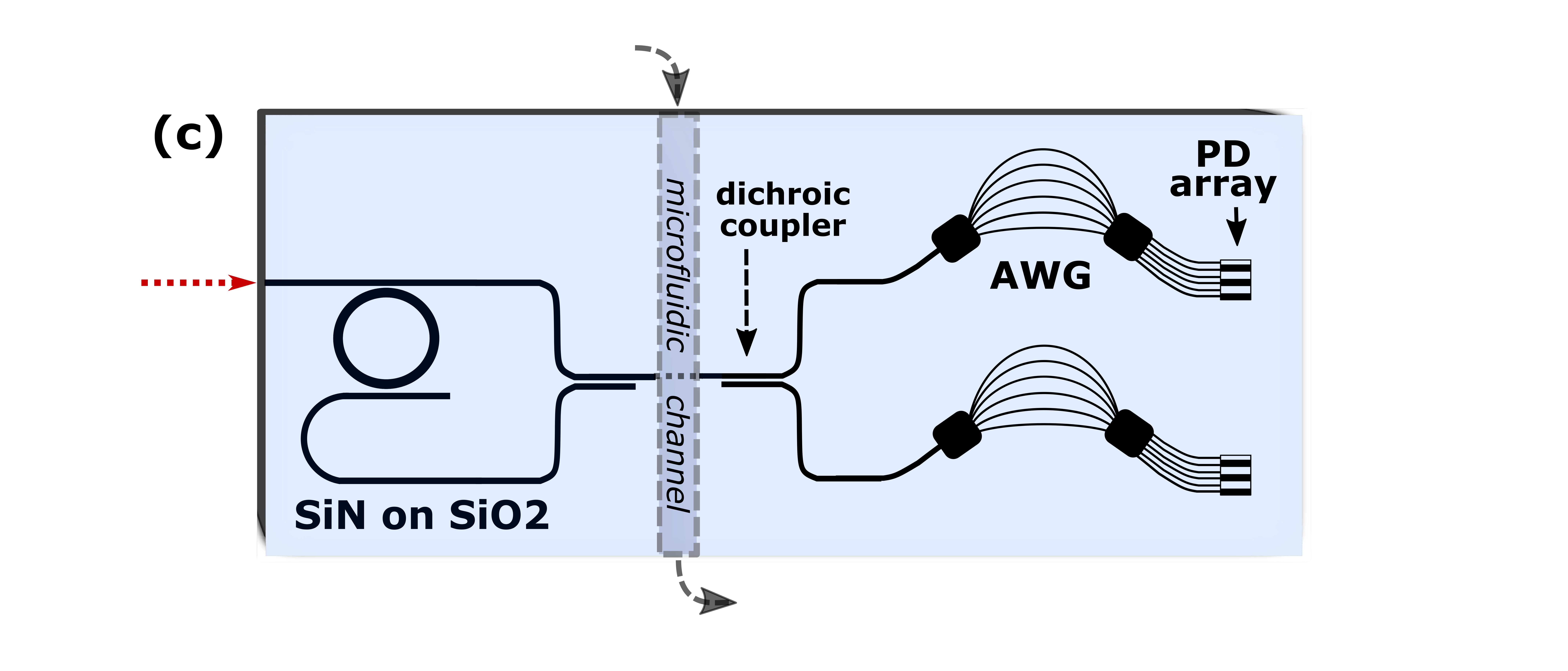}
\caption{Direct NIR spectroscopy of four liquid phase molecules using the SiN microcomb spectrum. (a) The full microcomb spectrum going through a blank cuvette (blue) is compared to the spectrum when sent through a cuvette containing a liquid of interest (red). (b) By dividing the two spectra as shown in part a, we find the absorbance of 1000-2300nm light through 1 cm of soybean oil, dichloromethane, or ethanol or through 1 mm of liquid water. Each (red) data point corresponds to a single SiN comb mode, and the solid black lines correspond to literature values for the absorption, as described in the text. Data points are connected by a red dashed line to guide the eye. (c) Notional drawing of a chip-scale spectrometer. The octave spanning microcomb spectrum is coupled from the chip into free space and sent through a sample. The spectrum minus the absorbed light is then coupled onto another chip containing a combination of adiabatic dichroic couplers and array waveguide gratings (AWG) such that individual comb modes are coupled into individual waveguides and sent to a photodetector (PD) array to measure the intensity per mode.} \label{fig:spectroscopy}
\end{figure}

To explore the potential usefulness of our microcomb setup for NIR spectroscopy applications, we perform direct spectroscopy on liquid water, soybean oil (lipid), dichloromethane, and ethanol (Fig. \ref{fig:spectroscopy}). The comb light is outcoupled from the chip and sent through an in-line fiber mount, where it passes through a cuvette containing the liquid, is collected by a multi-mode fiber, and is sent to an optical spectrum analyzer (OSA). The results are shown in figure \ref{fig:spectroscopy}.
The absorbance of each liquid is equal to the loss of intensity in a single microcomb mode as the light passes through a specified path length of the sample; we use an empty cuvette as the absorbance calibration. We compare to the absorption values measured in the literature, using the Beer-Lambert formula, $A=e^{-\alpha * PL}$, where A is the absorbance, $\alpha$ is the given absorption, and PL is the sample path length. The absorption of liquid water is taken from reference \cite{Palmer1974} (Figure \ref{fig:spectroscopy}b, solid black line in lower left plot), but quantitative values for the other three molecules in their liquid phases are hard to find. The solid red line in the upper left plot of figure \ref{fig:spectroscopy}b is calculated from the measured absorption of pig lard increased in concentration by a factor of 4 \cite{Veen2004}. Pig lard and soybean oil, though obviously different in origin, are both lipids with some general features in common due to their similar composition.

Figure \ref{fig:spectroscopy}c is a notional diagram describing how a microcomb might be incorporated into a chip-scale spectrometer. The chip contains a microresonator and waveguide coupler network on the left side, dichroic couplers \cite{Stanton2015} and arrayed waveguide gratings \cite{Gatkine2017,Stanton2017} on the right side, and the frequency comb light is directed through a microfluidic channel that carries a sample for study running down the center of the chip. The intensity in each waveguide would then be measured by a photodetector (PD). If absolute frequency calibration of the spectroscopy is useful, the f-2f approach and repetition frequency stabilization of the microcomb could be used as explored above in this paper. 

\vspace{-9pt}\section{Summary}
In conclusion, this work demonstrates the utility and generality of Kerr-microresonator optical frequency combs in a laboratory setting, as well as paths for further development towards chip-scale devices. OFD in a microscale device is a very promising direction for the eventual construction and commercialization of clock technology. Additionally, the flexibility provided by electro-optic modulation of the mode spacing provides microcomb devices with the capability to be used in many avenues of scientific research, from the many applications of spectroscopy, to further frequency-comb development, to massively parallel telecommunication.

\pagebreak

\part*{\centering{Supplementary Material}}

\section*{Control of $f_{ceo}$ using an auxiliary laser}

To overcome the inefficiency of our PPLN doubling process and the loss of light due to waveguide to fiber coupling, we employ a Newport Velocity TLB-6736 ($\lambda \approx 2$ microns) as an auxiliary laser ($\nu_{aux}$). The auxiliary laser light is sent through the PPLN, and 32 mW of light at 1998 nm and 1 mW of light at 999 nm are collected into the output fiber. This light is then use to measure two heterodynes ($\delta_{1}$ and $\delta_{2}$) with SiN comb modes (where $\nu_{f}$ is the optical frequency of $f$-th the microcomb mode at 1998 nm, and $\nu_{2f}$ is the frequency of the $(2f)$-th mode at 999 nm).
\begin{align*}
\delta_{1}&=\nu_{f}-\nu_{aux} \\
\delta_{2}&= \nu_{2f}-2\nu_{aux}
\end{align*}

Using the comb equation, we can show that doubling $\delta_{1}$ and mixing the two signals allows us to directly measure $f_{ceo}$.
\begin{align*}
\nu_{n} &= f_{ceo}+n*f_{rep} \\
2\delta_{1}-\delta_{2}&=2(\nu_{f}-\nu_{aux}) - (\nu_{2f}-2\nu_{aux}) \\
&=2(f_{ceo}+f*f_{rep})-(f_{ceo}+2f*f_{rep}) \\
&= f_{ceo}
\end{align*}

In the experiment, we digitally divide $\delta_{1}$ by 32 and $\delta_{2}$ by 64 before mixing. We then measure $f_{ceo}/64$ and control this signal using a PID feedback loop. As $|f_{ceo}| \approx 13$ GHz for this microresonator, the division factor makes the choice of electronics involved in the locking process more managable. One large improvement to the $f_{ceo}$ servo and the residual noise $f_{rep}$ shown in Fig. 2 of the main paper would come from choosing a microresonator with a lower $|f_{ceo}|$. Reference [19] outlines several ways to accomplish this.

\section*{Residual noise on THz mode spacing: experimental details}
\begin{figure}[htbp]
	\includegraphics[width=0.95\linewidth]{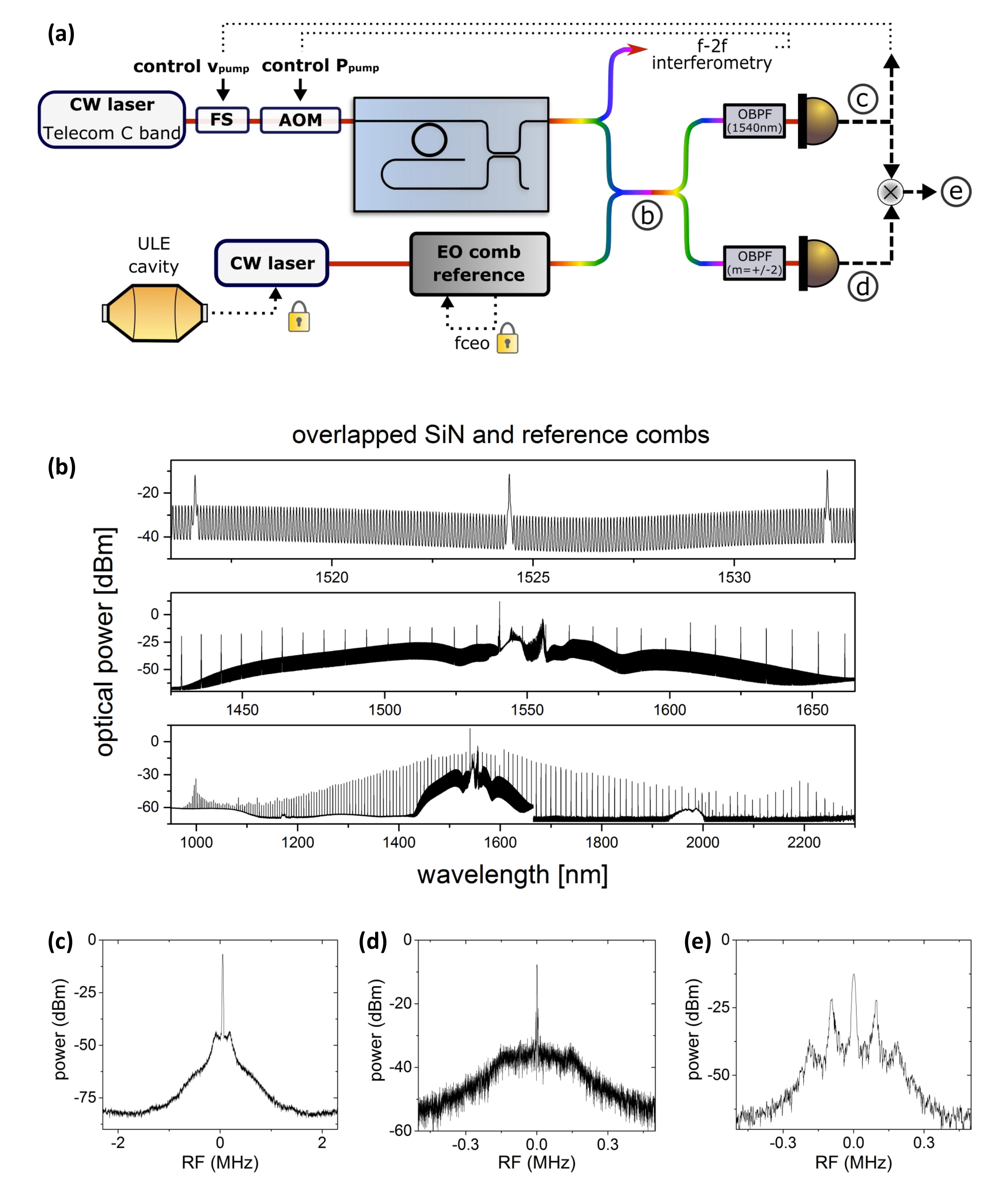}
	\caption{(a) Detailed setup diagram showing system stabilization and residual noise measurement using the reference EO comb. Letters refer to subsequent spectra and RF heterodyne signals. (b) Overlapped spectra of the SiN microcomb with THz mode spacing and the reference EO comb with 10 GHz mode spacing. (c) SiN pump (1540 nm) heterodyned against the closest EO comb mode. This beat is controlled via an electronic feedback circuit acting on the SiN pump frequency. (d) The SiN mode 2 THz away from the pump is heterodyned against the nearest EO comb mode. This signal is not electronically controlled. (e) The beats in (c) and (d) are mixed together, and the resulting beat measuring $f_{diff}$ is counted.}
	\label{fig:RNsetup}
\end{figure}

Figure \ref{fig:RNsetup} gives a detailed explanation of how the residual noise measurement is achieved. The signal shown in figure \ref{fig:RNsetup}e is measured with a phase noise analyzer and a counter, as seen in figure 2 of the main text. By mixing the signals in figure \ref{fig:RNsetup}c and \ref{fig:RNsetup}d, any effects of the servo circuit in \ref{fig:RNsetup}c are removed from the measurement such that the counted signal only reflects the residual noise on the mode spacings of either comb.

\newpage

\section*{Cavity drift cancellation in the residual noise measurement}

Due to the common mode nature of the residual noise measurement, most, though not all, of the drift of the cavity mode will be canceled in the measurement of $f_{diff}$.

Cavity drift, caused mainly by slow temperature drifts in the laboratory, is written onto both the repetition rate of the EO comb reference, $f_{rep EO}$, and the repetition rate of the SiN microcomb, $f_{rep THz}$.  The EO comb is seeded by a laser stabilized by the cavity, and $f_{rep EO}$ is dependent on the frequency of the cavity-stabilized laser, ${\nu}_{cavity}$, as follows:

\begin{align*}
{\nu}_{pump EO}&=f_{ceo EO}+M*f_{rep EO},  \\
f_{rep EO}&=\frac{{\nu}_{cavity}-f_{ceo EO}}{M},
\end{align*}

where $f_{ceo EO}$ is the carrier envelope offset frequency of the EO comb which is set by a local oscillator (LO) and PID servo loop, and $M$ is the mode number of the EO comb pump frequency, which is 19,339 for these experiments.

The frequency of the SiN microcomb pump is also stabilized by the cavity, but in this case we use a mode of the reference EO comb such in order to span the distance between the cavity mode at 1550 nm and the SiN resonator mode at 1540 nm:

\begin{align*}
{\nu}_{pump THz}&=f_{ceo THz}+N*f_{rep THz},  \\
f_{rep THz}&=\frac{(q*f_{rep EO}+{\nu}_{cavity})-f_{ceo THz}}{N},
\label{eq:refname1}
\end{align*}

where $f_{ceo THz}$ is the carrier envelope offset frequency set by an LO (described in section II of the main text and section I of this document), $N$ is the mode number of the SiN pump, which is 192 in these experiments, and $q$ is the number of EO comb modes between ${\nu}_{cavity}$ and ${\nu}_{pump THz}$.  Thus, we see that the repetition rates of both combs depend on ${\nu}_{cavity}$ with differing rates of dependence.

Our measurement of the residual noise on the THz repetition rate takes into account this small dependence on the cavity drift.  We measure the electronically mixed difference of two heterodyne beats between the two combs, $f_{diff}$:
\begin{align*}
f_{diff}&=(\nu_{1 EO}-\nu_{1 THz})-(\nu_{2 EO}-\nu_{2 THz}),  \\
f_{diff}&=A*f_{rep EO}-B*f_{rep THz},
\end{align*}

where $A$ and $B$ are the number of EO and SiN modes separating the two heterodynes at 1540nm and 1524nm.  (For this experiment, $A=203$ and $B=2$.)  Thus, $f_{diff}$ has a dependence on $\nu_{cavity}$ such that:

\begin{gather*}
f_{diff}=A*\frac{{\nu}_{cavity}-f_{ceo EO}}{M}\\-B*\frac{(\frac{q}{M}({\nu}_{cavity}-f_{ceo EO})+{\nu}_{cavity}-f_{ceo THz}}{N} \\
\frac{d f_{diff}}{d {\nu}_{cavity}}=\frac{A}{M}-\frac{B}{N}*(1+\frac{q}{M})
\end{gather*}

and for this experiment, $\frac{d f_{diff}}{d {\nu}_{cavity}} \ $$\approx$$ \ 1.5*10^{-5}$. 

\begin{figure}[htbp]
	\centering
	\includegraphics[width=\linewidth]{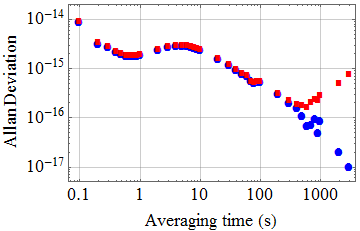}
	\caption{Correcting for cavity drift in residual noise measurement. Red squares reflect the Allan deviation of the raw data, while the blue circles show the Allan deviation after a linear drift of 47 mHz per second, originating from the cavity and measured separately, is accounted for.}
	\label{fig:Adevcavitydrift}
\end{figure}

Figure \ref{fig:Adevcavitydrift} shows the Allan deviation of the residual noise beat before cavity drift is taken into account.  Note the small upturn for averaging times above 100 seconds.  By taking a simultaneous measurement of the EO comb repetition rate, we are able to obtain a real time measurement of the cavity drift, which can then be subtracted from $f_{diff}$.  We found the cavity drift to be 47 mHz/s during this 2 hour period, and figure \ref{fig:Adevcavitydrift} in the main text reflects the Allan deviation after removing this drift.

\vspace{-9pt}\section{Acknowledgments}
The authors thank Srico, Inc. for use of the waveguide PPLN device used in f-2f interferometery. We thank Martin Pfeiffer and Tobias Kippenberg for the use of the auxiliary laser; Wei Zhang, Josue Davila-Rodriguez, Kevin Cossel, Gabriel Ycas, Bill Swann, and Nate Newbury for the use of cavity-stabilized laser light; Eleanor Waxman and David Plusquellic for advice regarding NIR spectroscopy; Susan Schima and Sonja Ringen for help preparing spectroscopic samples; and the DARPA DODOS team. This research is supported by the Defense Advanced Research Projects Agency DODOS and ACES programs, NRC, AFOSR (FA9550-16-1-0016), NASA, and NIST. This work is not subject to copyright in the United States.

\bibliography{OFDpaperfixed} 

\begin{thebibliography}{44}%
\makeatletter
\providecommand \@ifxundefined [1]{%
 \@ifx{#1\undefined}
}%
\providecommand \@ifnum [1]{%
 \ifnum #1\expandafter \@firstoftwo
 \else \expandafter \@secondoftwo
 \fi
}%
\providecommand \@ifx [1]{%
 \ifx #1\expandafter \@firstoftwo
 \else \expandafter \@secondoftwo
 \fi
}%
\providecommand \natexlab [1]{#1}%
\providecommand \enquote  [1]{``#1''}%
\providecommand \bibnamefont  [1]{#1}%
\providecommand \bibfnamefont [1]{#1}%
\providecommand \citenamefont [1]{#1}%
\providecommand \href@noop [0]{\@secondoftwo}%
\providecommand \href [0]{\begingroup \@sanitize@url \@href}%
\providecommand \@href[1]{\@@startlink{#1}\@@href}%
\providecommand \@@href[1]{\endgroup#1\@@endlink}%
\providecommand \@sanitize@url [0]{\catcode `\\12\catcode `\$12\catcode
  `\&12\catcode `\#12\catcode `\^12\catcode `\_12\catcode `\%12\relax}%
\providecommand \@@startlink[1]{}%
\providecommand \@@endlink[0]{}%
\providecommand \url  [0]{\begingroup\@sanitize@url \@url }%
\providecommand \@url [1]{\endgroup\@href {#1}{\urlprefix }}%
\providecommand \urlprefix  [0]{URL }%
\providecommand \Eprint [0]{\href }%
\providecommand \doibase [0]{http://dx.doi.org/}%
\providecommand \selectlanguage [0]{\@gobble}%
\providecommand \bibinfo  [0]{\@secondoftwo}%
\providecommand \bibfield  [0]{\@secondoftwo}%
\providecommand \translation [1]{[#1]}%
\providecommand \BibitemOpen [0]{}%
\providecommand \bibitemStop [0]{}%
\providecommand \bibitemNoStop [0]{.\EOS\space}%
\providecommand \EOS [0]{\spacefactor3000\relax}%
\providecommand \BibitemShut  [1]{\csname bibitem#1\endcsname}%
\let\auto@bib@innerbib\@empty
\bibitem [{\citenamefont {Ludlow}\ \emph {et~al.}(2015)\citenamefont {Ludlow},
  \citenamefont {Boyd}, \citenamefont {Ye}, \citenamefont {Peik},\ and\
  \citenamefont {Schmidt}}]{Ludlow2015}%
  \BibitemOpen
  \bibfield  {author} {\bibinfo {author} {\bibfnamefont {A.~D.}\ \bibnamefont
  {Ludlow}}, \bibinfo {author} {\bibfnamefont {M.~M.}\ \bibnamefont {Boyd}},
  \bibinfo {author} {\bibfnamefont {J.}~\bibnamefont {Ye}}, \bibinfo {author}
  {\bibfnamefont {E.}~\bibnamefont {Peik}}, \ and\ \bibinfo {author}
  {\bibfnamefont {P.~O.}\ \bibnamefont {Schmidt}},\ }\href {\doibase
  10.1103/RevModPhys.87.637} {\bibfield  {journal} {\bibinfo  {journal}
  {Reviews of Modern Physics}\ }\textbf {\bibinfo {volume} {87}},\ \bibinfo
  {pages} {637} (\bibinfo {year} {2015})}\BibitemShut {NoStop}%
\bibitem [{\citenamefont {Diddams}\ \emph {et~al.}(2001)\citenamefont
  {Diddams}, \citenamefont {Udem}, \citenamefont {Bergquist}, \citenamefont
  {Curtis}, \citenamefont {Drullinger}, \citenamefont {Hollberg}, \citenamefont
  {Itano}, \citenamefont {Lee}, \citenamefont {Oates}, \citenamefont {Vogel},\
  and\ \citenamefont {Wineland}}]{Diddams2001}%
  \BibitemOpen
  \bibfield  {author} {\bibinfo {author} {\bibfnamefont {S.~A.}\ \bibnamefont
  {Diddams}}, \bibinfo {author} {\bibfnamefont {T.}~\bibnamefont {Udem}},
  \bibinfo {author} {\bibfnamefont {J.~C.}\ \bibnamefont {Bergquist}}, \bibinfo
  {author} {\bibfnamefont {E.~A.}\ \bibnamefont {Curtis}}, \bibinfo {author}
  {\bibfnamefont {R.~E.}\ \bibnamefont {Drullinger}}, \bibinfo {author}
  {\bibfnamefont {L.}~\bibnamefont {Hollberg}}, \bibinfo {author}
  {\bibfnamefont {W.~M.}\ \bibnamefont {Itano}}, \bibinfo {author}
  {\bibfnamefont {W.~D.}\ \bibnamefont {Lee}}, \bibinfo {author} {\bibfnamefont
  {C.~W.}\ \bibnamefont {Oates}}, \bibinfo {author} {\bibfnamefont {K.~R.}\
  \bibnamefont {Vogel}}, \ and\ \bibinfo {author} {\bibfnamefont {D.~J.}\
  \bibnamefont {Wineland}},\ }\href {\doibase 10.1126/science.1061171}
  {\bibfield  {journal} {\bibinfo  {journal} {Science}\ }\textbf {\bibinfo
  {volume} {293}},\ \bibinfo {pages} {825} (\bibinfo {year}
  {2001})}\BibitemShut {NoStop}%
\bibitem [{\citenamefont {Ye}\ \emph {et~al.}(2000)\citenamefont {Ye},
  \citenamefont {Hall},\ and\ \citenamefont {Diddams}}]{Diddams2000_1}%
  \BibitemOpen
  \bibfield  {author} {\bibinfo {author} {\bibfnamefont {J.}~\bibnamefont
  {Ye}}, \bibinfo {author} {\bibfnamefont {J.~L.}\ \bibnamefont {Hall}}, \ and\
  \bibinfo {author} {\bibfnamefont {S.~A.}\ \bibnamefont {Diddams}},\
  }\href@noop {} {\bibfield  {journal} {\bibinfo  {journal} {Optics Letters}\
  }\textbf {\bibinfo {volume} {25}},\ \bibinfo {pages} {1675} (\bibinfo {year}
  {2000})}\BibitemShut {NoStop}%
\bibitem [{\citenamefont {Bartels}\ \emph {et~al.}(2005)\citenamefont
  {Bartels}, \citenamefont {Diddams}, \citenamefont {Oates}, \citenamefont
  {Wilpers}, \citenamefont {Bergquist}, \citenamefont {Oskay},\ and\
  \citenamefont {Hollberg}}]{Bartels2005}%
  \BibitemOpen
  \bibfield  {author} {\bibinfo {author} {\bibfnamefont {A.}~\bibnamefont
  {Bartels}}, \bibinfo {author} {\bibfnamefont {S.~A.}\ \bibnamefont
  {Diddams}}, \bibinfo {author} {\bibfnamefont {C.~W.}\ \bibnamefont {Oates}},
  \bibinfo {author} {\bibfnamefont {G.}~\bibnamefont {Wilpers}}, \bibinfo
  {author} {\bibfnamefont {J.~C.}\ \bibnamefont {Bergquist}}, \bibinfo {author}
  {\bibfnamefont {W.~H.}\ \bibnamefont {Oskay}}, \ and\ \bibinfo {author}
  {\bibfnamefont {L.}~\bibnamefont {Hollberg}},\ }\href {\doibase
  10.1364/OL.30.000667} {\bibfield  {journal} {\bibinfo  {journal} {Optics
  Letters}\ }\textbf {\bibinfo {volume} {30}},\ \bibinfo {pages} {667}
  (\bibinfo {year} {2005})}\BibitemShut {NoStop}%
\bibitem [{\citenamefont {McFerran}\ \emph {et~al.}(2005)\citenamefont
  {McFerran}, \citenamefont {Ivanov}, \citenamefont {Bartels}, \citenamefont
  {Wilpers}, \citenamefont {Oates}, \citenamefont {Diddams},\ and\
  \citenamefont {Hollberg}}]{McFerran2005}%
  \BibitemOpen
  \bibfield  {author} {\bibinfo {author} {\bibfnamefont {J.~J.}\ \bibnamefont
  {McFerran}}, \bibinfo {author} {\bibfnamefont {E.~N.}\ \bibnamefont
  {Ivanov}}, \bibinfo {author} {\bibfnamefont {A.}~\bibnamefont {Bartels}},
  \bibinfo {author} {\bibfnamefont {G.}~\bibnamefont {Wilpers}}, \bibinfo
  {author} {\bibfnamefont {C.~W.}\ \bibnamefont {Oates}}, \bibinfo {author}
  {\bibfnamefont {S.~A.}\ \bibnamefont {Diddams}}, \ and\ \bibinfo {author}
  {\bibfnamefont {L.}~\bibnamefont {Hollberg}},\ }\href {\doibase
  10.1049/el:20050505} {\bibfield  {journal} {\bibinfo  {journal} {Electronics
  Letters}\ }\textbf {\bibinfo {volume} {41}},\ \bibinfo {pages} {650}
  (\bibinfo {year} {2005})},\ \Eprint {http://arxiv.org/abs/0504102} {0504102}
  \BibitemShut {NoStop}%
\bibitem [{\citenamefont {Millo}\ \emph {et~al.}(2009)\citenamefont {Millo},
  \citenamefont {Abgrall}, \citenamefont {Lours}, \citenamefont {English},
  \citenamefont {Jiang}, \citenamefont {G{\'{u}}na}, \citenamefont {Clairon},
  \citenamefont {Tobar}, \citenamefont {Bize}, \citenamefont {{Le Coq}},\ and\
  \citenamefont {Santarelli}}]{Millo2009}%
  \BibitemOpen
  \bibfield  {author} {\bibinfo {author} {\bibfnamefont {J.}~\bibnamefont
  {Millo}}, \bibinfo {author} {\bibfnamefont {M.}~\bibnamefont {Abgrall}},
  \bibinfo {author} {\bibfnamefont {M.}~\bibnamefont {Lours}}, \bibinfo
  {author} {\bibfnamefont {E.~M.}\ \bibnamefont {English}}, \bibinfo {author}
  {\bibfnamefont {H.}~\bibnamefont {Jiang}}, \bibinfo {author} {\bibfnamefont
  {J.}~\bibnamefont {G{\'{u}}na}}, \bibinfo {author} {\bibfnamefont
  {A.}~\bibnamefont {Clairon}}, \bibinfo {author} {\bibfnamefont {M.~E.}\
  \bibnamefont {Tobar}}, \bibinfo {author} {\bibfnamefont {S.}~\bibnamefont
  {Bize}}, \bibinfo {author} {\bibfnamefont {Y.}~\bibnamefont {{Le Coq}}}, \
  and\ \bibinfo {author} {\bibfnamefont {G.}~\bibnamefont {Santarelli}},\
  }\href@noop {} {\bibfield  {journal} {\bibinfo  {journal} {Applied Physics
  Letters}\ }\textbf {\bibinfo {volume} {94}} (\bibinfo {year}
  {2009})}\BibitemShut {NoStop}%
\bibitem [{\citenamefont {Weyers}\ \emph {et~al.}(2009)\citenamefont {Weyers},
  \citenamefont {Lipphardt},\ and\ \citenamefont {Schnatz}}]{Weyers2009}%
  \BibitemOpen
  \bibfield  {author} {\bibinfo {author} {\bibfnamefont {S.}~\bibnamefont
  {Weyers}}, \bibinfo {author} {\bibfnamefont {B.}~\bibnamefont {Lipphardt}}, \
  and\ \bibinfo {author} {\bibfnamefont {H.}~\bibnamefont {Schnatz}},\
  }\href@noop {} {\bibfield  {journal} {\bibinfo  {journal} {Phys. Rev. A}\
  }\textbf {\bibinfo {volume} {79}} (\bibinfo {year} {2009})}\BibitemShut
  {NoStop}%
\bibitem [{\citenamefont {Zhang}\ \emph {et~al.}(2010)\citenamefont {Zhang},
  \citenamefont {Xu}, \citenamefont {Lours}, \citenamefont {Boudot},
  \citenamefont {Kersaĺ}, \citenamefont {Santarelli},\ and\ \citenamefont {{Le
  Coq}}}]{Zhang2010}%
  \BibitemOpen
  \bibfield  {author} {\bibinfo {author} {\bibfnamefont {W.}~\bibnamefont
  {Zhang}}, \bibinfo {author} {\bibfnamefont {Z.}~\bibnamefont {Xu}}, \bibinfo
  {author} {\bibfnamefont {M.}~\bibnamefont {Lours}}, \bibinfo {author}
  {\bibfnamefont {R.}~\bibnamefont {Boudot}}, \bibinfo {author} {\bibfnamefont
  {Y.}~\bibnamefont {Kersaĺ}}, \bibinfo {author} {\bibfnamefont
  {G.}~\bibnamefont {Santarelli}}, \ and\ \bibinfo {author} {\bibfnamefont
  {Y.}~\bibnamefont {{Le Coq}}},\ }\href@noop {} {\bibfield  {journal}
  {\bibinfo  {journal} {Applied Physics Letters}\ }\textbf {\bibinfo {volume}
  {96}} (\bibinfo {year} {2010})}\BibitemShut {NoStop}%
\bibitem [{\citenamefont {Fortier}\ \emph {et~al.}(2011)\citenamefont
  {Fortier}, \citenamefont {Kirchner}, \citenamefont {Quinlan}, \citenamefont
  {Taylor}, \citenamefont {Bergquist}, \citenamefont {Rosenband}, \citenamefont
  {Lemke}, \citenamefont {Ludlow}, \citenamefont {Jiang}, \citenamefont
  {Oates},\ and\ \citenamefont {Diddams}}]{Fortier2011}%
  \BibitemOpen
  \bibfield  {author} {\bibinfo {author} {\bibfnamefont {T.~M.}\ \bibnamefont
  {Fortier}}, \bibinfo {author} {\bibfnamefont {M.~S.}\ \bibnamefont
  {Kirchner}}, \bibinfo {author} {\bibfnamefont {F.}~\bibnamefont {Quinlan}},
  \bibinfo {author} {\bibfnamefont {J.}~\bibnamefont {Taylor}}, \bibinfo
  {author} {\bibfnamefont {J.~C.}\ \bibnamefont {Bergquist}}, \bibinfo {author}
  {\bibfnamefont {T.}~\bibnamefont {Rosenband}}, \bibinfo {author}
  {\bibfnamefont {N.}~\bibnamefont {Lemke}}, \bibinfo {author} {\bibfnamefont
  {A.}~\bibnamefont {Ludlow}}, \bibinfo {author} {\bibfnamefont
  {Y.}~\bibnamefont {Jiang}}, \bibinfo {author} {\bibfnamefont {C.~W.}\
  \bibnamefont {Oates}}, \ and\ \bibinfo {author} {\bibfnamefont {S.~A.}\
  \bibnamefont {Diddams}},\ }\href {\doibase 10.1038/nphoton.2011.121}
  {\bibfield  {journal} {\bibinfo  {journal} {Nature Photonics}\ }\textbf
  {\bibinfo {volume} {5}},\ \bibinfo {pages} {425} (\bibinfo {year}
  {2011})}\BibitemShut {NoStop}%
\bibitem [{\citenamefont {Baynes}\ \emph {et~al.}(2015)\citenamefont {Baynes},
  \citenamefont {Quinlan}, \citenamefont {Fortier}, \citenamefont {Zhou},
  \citenamefont {Beling}, \citenamefont {Campbell},\ and\ \citenamefont
  {Diddams}}]{Baynes2015}%
  \BibitemOpen
  \bibfield  {author} {\bibinfo {author} {\bibfnamefont {F.~N.}\ \bibnamefont
  {Baynes}}, \bibinfo {author} {\bibfnamefont {F.}~\bibnamefont {Quinlan}},
  \bibinfo {author} {\bibfnamefont {T.~M.}\ \bibnamefont {Fortier}}, \bibinfo
  {author} {\bibfnamefont {Q.}~\bibnamefont {Zhou}}, \bibinfo {author}
  {\bibfnamefont {A.}~\bibnamefont {Beling}}, \bibinfo {author} {\bibfnamefont
  {J.~C.}\ \bibnamefont {Campbell}}, \ and\ \bibinfo {author} {\bibfnamefont
  {S.~A.}\ \bibnamefont {Diddams}},\ }\href {\doibase 10.1364/OPTICA.2.000141}
  {\bibfield  {journal} {\bibinfo  {journal} {Optica}\ }\textbf {\bibinfo
  {volume} {2}},\ \bibinfo {pages} {141} (\bibinfo {year} {2015})}\BibitemShut
  {NoStop}%
\bibitem [{\citenamefont {Xie}\ \emph {et~al.}(2016)\citenamefont {Xie},
  \citenamefont {Bouchand}, \citenamefont {Nicolodi}, \citenamefont {Giunta},
  \citenamefont {Hänsel}, \citenamefont {Lezius}, \citenamefont {Joshi},
  \citenamefont {Datta}, \citenamefont {Alexandre}, \citenamefont {Lours},
  \citenamefont {Tremblin}, \citenamefont {Santarelli}, \citenamefont
  {Holzwarth},\ and\ \citenamefont {Le~Coq}}]{Xie2016}%
  \BibitemOpen
  \bibfield  {author} {\bibinfo {author} {\bibfnamefont {X.}~\bibnamefont
  {Xie}}, \bibinfo {author} {\bibfnamefont {R.}~\bibnamefont {Bouchand}},
  \bibinfo {author} {\bibfnamefont {D.}~\bibnamefont {Nicolodi}}, \bibinfo
  {author} {\bibfnamefont {M.}~\bibnamefont {Giunta}}, \bibinfo {author}
  {\bibfnamefont {W.}~\bibnamefont {Hänsel}}, \bibinfo {author} {\bibfnamefont
  {M.}~\bibnamefont {Lezius}}, \bibinfo {author} {\bibfnamefont
  {A.}~\bibnamefont {Joshi}}, \bibinfo {author} {\bibfnamefont
  {S.}~\bibnamefont {Datta}}, \bibinfo {author} {\bibfnamefont
  {C.}~\bibnamefont {Alexandre}}, \bibinfo {author} {\bibfnamefont
  {M.}~\bibnamefont {Lours}}, \bibinfo {author} {\bibfnamefont {P.-A.}\
  \bibnamefont {Tremblin}}, \bibinfo {author} {\bibfnamefont {G.}~\bibnamefont
  {Santarelli}}, \bibinfo {author} {\bibfnamefont {R.}~\bibnamefont
  {Holzwarth}}, \ and\ \bibinfo {author} {\bibfnamefont {Y.}~\bibnamefont
  {Le~Coq}},\ }\href@noop {} {\bibfield  {journal} {\bibinfo  {journal} {Nature
  Photonics}\ }\textbf {\bibinfo {volume} {11}},\ \bibinfo {pages} {44}
  (\bibinfo {year} {2016})}\BibitemShut {NoStop}%
\bibitem [{\citenamefont {Kippenberg}\ \emph {et~al.}(2018)\citenamefont
  {Kippenberg}, \citenamefont {Gaeta}, \citenamefont {Lipson},\ and\
  \citenamefont {Gorodetsky}}]{Kippenbergeaan8083}%
  \BibitemOpen
  \bibfield  {author} {\bibinfo {author} {\bibfnamefont {T.~J.}\ \bibnamefont
  {Kippenberg}}, \bibinfo {author} {\bibfnamefont {A.~L.}\ \bibnamefont
  {Gaeta}}, \bibinfo {author} {\bibfnamefont {M.}~\bibnamefont {Lipson}}, \
  and\ \bibinfo {author} {\bibfnamefont {M.~L.}\ \bibnamefont {Gorodetsky}},\
  }\href@noop {} {\bibfield  {journal} {\bibinfo  {journal} {Science}\ }\textbf
  {\bibinfo {volume} {361}} (\bibinfo {year} {2018})}\BibitemShut {NoStop}%
\bibitem [{\citenamefont {Agrawal}(2013)}]{AgrawalBook}%
  \BibitemOpen
  \bibfield  {author} {\bibinfo {author} {\bibfnamefont {G.~P.}\ \bibnamefont
  {Agrawal}},\ }\href@noop {} {\emph {\bibinfo {title} {Nonlinear Fiber
  Optics}}},\ \bibinfo {edition} {5th}\ ed.\ (\bibinfo  {publisher} {Academic
  Press},\ \bibinfo {year} {2013})\BibitemShut {NoStop}%
\bibitem [{\citenamefont {Cole}\ \emph {et~al.}(2017)\citenamefont {Cole},
  \citenamefont {Lamb}, \citenamefont {Del'Haye}, \citenamefont {Diddams},\
  and\ \citenamefont {Papp}}]{Cole2017}%
  \BibitemOpen
  \bibfield  {author} {\bibinfo {author} {\bibfnamefont {D.~C.}\ \bibnamefont
  {Cole}}, \bibinfo {author} {\bibfnamefont {E.~S.}\ \bibnamefont {Lamb}},
  \bibinfo {author} {\bibfnamefont {P.}~\bibnamefont {Del'Haye}}, \bibinfo
  {author} {\bibfnamefont {S.~A.}\ \bibnamefont {Diddams}}, \ and\ \bibinfo
  {author} {\bibfnamefont {S.~B.}\ \bibnamefont {Papp}},\ }\href {\doibase
  10.1038/s41566-017-0009-z} {\bibfield  {journal} {\bibinfo  {journal} {Nature
  Photonics}\ }\textbf {\bibinfo {volume} {11}},\ \bibinfo {pages} {671}
  (\bibinfo {year} {2017})}\BibitemShut {NoStop}%
\bibitem [{\citenamefont {Yao}\ \emph {et~al.}(2018)\citenamefont {Yao},
  \citenamefont {Huang}, \citenamefont {Liu}, \citenamefont {Vinod},
  \citenamefont {Choi}, \citenamefont {Hoff}, \citenamefont {Li}, \citenamefont
  {Yu}, \citenamefont {Feng}, \citenamefont {Kwong}, \citenamefont {Huang},
  \citenamefont {Rao}, \citenamefont {Duan},\ and\ \citenamefont
  {Wong}}]{Yao2018}%
  \BibitemOpen
  \bibfield  {author} {\bibinfo {author} {\bibfnamefont {B.}~\bibnamefont
  {Yao}}, \bibinfo {author} {\bibfnamefont {S.-W.}\ \bibnamefont {Huang}},
  \bibinfo {author} {\bibfnamefont {Y.}~\bibnamefont {Liu}}, \bibinfo {author}
  {\bibfnamefont {A.~K.}\ \bibnamefont {Vinod}}, \bibinfo {author}
  {\bibfnamefont {C.}~\bibnamefont {Choi}}, \bibinfo {author} {\bibfnamefont
  {M.}~\bibnamefont {Hoff}}, \bibinfo {author} {\bibfnamefont {Y.}~\bibnamefont
  {Li}}, \bibinfo {author} {\bibfnamefont {M.}~\bibnamefont {Yu}}, \bibinfo
  {author} {\bibfnamefont {Z.}~\bibnamefont {Feng}}, \bibinfo {author}
  {\bibfnamefont {D.-L.}\ \bibnamefont {Kwong}}, \bibinfo {author}
  {\bibfnamefont {Y.}~\bibnamefont {Huang}}, \bibinfo {author} {\bibfnamefont
  {Y.}~\bibnamefont {Rao}}, \bibinfo {author} {\bibfnamefont {X.}~\bibnamefont
  {Duan}}, \ and\ \bibinfo {author} {\bibfnamefont {C.~W.}\ \bibnamefont
  {Wong}},\ }\href {\doibase 10.1038/s41586-018-0216-x} {\bibfield  {journal}
  {\bibinfo  {journal} {Nature}\ }\textbf {\bibinfo {volume} {558}},\ \bibinfo
  {pages} {410–414} (\bibinfo {year} {2018})}\BibitemShut {NoStop}%
\bibitem [{\citenamefont {Li}\ \emph {et~al.}(2016)\citenamefont {Li},
  \citenamefont {Briles}, \citenamefont {Westly}, \citenamefont {Drake},
  \citenamefont {Stone}, \citenamefont {Ilic}, \citenamefont {Diddams},
  \citenamefont {Papp},\ and\ \citenamefont {Srinivasan}}]{Li2016}%
  \BibitemOpen
  \bibfield  {author} {\bibinfo {author} {\bibfnamefont {Q.}~\bibnamefont
  {Li}}, \bibinfo {author} {\bibfnamefont {T.~C.}\ \bibnamefont {Briles}},
  \bibinfo {author} {\bibfnamefont {D.~A.}\ \bibnamefont {Westly}}, \bibinfo
  {author} {\bibfnamefont {T.~E.}\ \bibnamefont {Drake}}, \bibinfo {author}
  {\bibfnamefont {J.~R.}\ \bibnamefont {Stone}}, \bibinfo {author}
  {\bibfnamefont {B.~R.}\ \bibnamefont {Ilic}}, \bibinfo {author}
  {\bibfnamefont {S.~A.}\ \bibnamefont {Diddams}}, \bibinfo {author}
  {\bibfnamefont {S.~B.}\ \bibnamefont {Papp}}, \ and\ \bibinfo {author}
  {\bibfnamefont {K.}~\bibnamefont {Srinivasan}},\ }\href
  {http://arxiv.org/abs/1611.09229 http://dx.doi.org/10.1364/OPTICA.4.000193}
  {\bibfield  {journal} {\bibinfo  {journal} {Optica}\ }\textbf {\bibinfo
  {volume} {4}} (\bibinfo {year} {2016})}\BibitemShut {NoStop}%
\bibitem [{\citenamefont {Yi}\ \emph {et~al.}(2015)\citenamefont {Yi},
  \citenamefont {Yang}, \citenamefont {Yang}, \citenamefont {Suh},\ and\
  \citenamefont {Vahala}}]{Yi2015}%
  \BibitemOpen
  \bibfield  {author} {\bibinfo {author} {\bibfnamefont {X.}~\bibnamefont
  {Yi}}, \bibinfo {author} {\bibfnamefont {Q.-F.}\ \bibnamefont {Yang}},
  \bibinfo {author} {\bibfnamefont {K.~Y.}\ \bibnamefont {Yang}}, \bibinfo
  {author} {\bibfnamefont {M.-G.}\ \bibnamefont {Suh}}, \ and\ \bibinfo
  {author} {\bibfnamefont {K.}~\bibnamefont {Vahala}},\ }\href {\doibase
  10.1364/OPTICA.2.001078} {\bibfield  {journal} {\bibinfo  {journal} {Optica}\
  }\textbf {\bibinfo {volume} {2}},\ \bibinfo {pages} {1078} (\bibinfo {year}
  {2015})}\BibitemShut {NoStop}%
\bibitem [{\citenamefont {Papp}\ \emph {et~al.}(2014)\citenamefont {Papp},
  \citenamefont {Beha}, \citenamefont {Del'Haye}, \citenamefont {Quinlan},
  \citenamefont {Lee}, \citenamefont {Vahala},\ and\ \citenamefont
  {Diddams}}]{Papp2014}%
  \BibitemOpen
  \bibfield  {author} {\bibinfo {author} {\bibfnamefont {S.~B.}\ \bibnamefont
  {Papp}}, \bibinfo {author} {\bibfnamefont {K.}~\bibnamefont {Beha}}, \bibinfo
  {author} {\bibfnamefont {P.}~\bibnamefont {Del'Haye}}, \bibinfo {author}
  {\bibfnamefont {F.}~\bibnamefont {Quinlan}}, \bibinfo {author} {\bibfnamefont
  {H.}~\bibnamefont {Lee}}, \bibinfo {author} {\bibfnamefont {K.~J.}\
  \bibnamefont {Vahala}}, \ and\ \bibinfo {author} {\bibfnamefont {S.~A.}\
  \bibnamefont {Diddams}},\ }\href {\doibase 10.1364/OPTICA.1.000010}
  {\bibfield  {journal} {\bibinfo  {journal} {Optica}\ }\textbf {\bibinfo
  {volume} {1}},\ \bibinfo {pages} {10} (\bibinfo {year} {2014})}\BibitemShut
  {NoStop}%
\bibitem [{\citenamefont {Briles}\ \emph {et~al.}(2018)\citenamefont {Briles},
  \citenamefont {Stone}, \citenamefont {Drake}, \citenamefont {Spencer},
  \citenamefont {Fredrick}, \citenamefont {Li}, \citenamefont {Westly},
  \citenamefont {Ilic}, \citenamefont {Srinivasan}, \citenamefont {Diddams},\
  and\ \citenamefont {Papp}}]{Briles2017}%
  \BibitemOpen
  \bibfield  {author} {\bibinfo {author} {\bibfnamefont {T.~C.}\ \bibnamefont
  {Briles}}, \bibinfo {author} {\bibfnamefont {J.~R.}\ \bibnamefont {Stone}},
  \bibinfo {author} {\bibfnamefont {T.~E.}\ \bibnamefont {Drake}}, \bibinfo
  {author} {\bibfnamefont {D.~T.}\ \bibnamefont {Spencer}}, \bibinfo {author}
  {\bibfnamefont {C.}~\bibnamefont {Fredrick}}, \bibinfo {author}
  {\bibfnamefont {Q.}~\bibnamefont {Li}}, \bibinfo {author} {\bibfnamefont
  {D.}~\bibnamefont {Westly}}, \bibinfo {author} {\bibfnamefont {B.~R.}\
  \bibnamefont {Ilic}}, \bibinfo {author} {\bibfnamefont {K.}~\bibnamefont
  {Srinivasan}}, \bibinfo {author} {\bibfnamefont {S.~A.}\ \bibnamefont
  {Diddams}}, \ and\ \bibinfo {author} {\bibfnamefont {S.~B.}\ \bibnamefont
  {Papp}},\ }\href {\doibase 10.1364/OL.43.002933} {\bibfield  {journal}
  {\bibinfo  {journal} {Opt. Lett.}\ }\textbf {\bibinfo {volume} {43}},\
  \bibinfo {pages} {2933} (\bibinfo {year} {2018})}\BibitemShut {NoStop}%
\bibitem [{\citenamefont {Spencer}\ \emph {et~al.}(2018)\citenamefont
  {Spencer}, \citenamefont {Drake}, \citenamefont {Briles}, \citenamefont
  {Stone}, \citenamefont {Sinclair}, \citenamefont {Fredrick}, \citenamefont
  {Li}, \citenamefont {Westly}, \citenamefont {Ilic}, \citenamefont
  {Bluestone}, \citenamefont {Volet}, \citenamefont {Komljenovic},
  \citenamefont {Chang}, \citenamefont {Lee}, \citenamefont {Oh}, \citenamefont
  {Suh}, \citenamefont {Yang}, \citenamefont {Pfeiffer}, \citenamefont
  {Kippenberg}, \citenamefont {Norberg}, \citenamefont {Theogarajan},
  \citenamefont {Vahala}, \citenamefont {Newbury}, \citenamefont {Srinivasan},
  \citenamefont {Bowers}, \citenamefont {Diddams},\ and\ \citenamefont
  {Papp}}]{Spencer2017}%
  \BibitemOpen
  \bibfield  {author} {\bibinfo {author} {\bibfnamefont {D.~T.}\ \bibnamefont
  {Spencer}}, \bibinfo {author} {\bibfnamefont {T.}~\bibnamefont {Drake}},
  \bibinfo {author} {\bibfnamefont {T.~C.}\ \bibnamefont {Briles}}, \bibinfo
  {author} {\bibfnamefont {J.}~\bibnamefont {Stone}}, \bibinfo {author}
  {\bibfnamefont {L.~C.}\ \bibnamefont {Sinclair}}, \bibinfo {author}
  {\bibfnamefont {C.}~\bibnamefont {Fredrick}}, \bibinfo {author}
  {\bibfnamefont {Q.}~\bibnamefont {Li}}, \bibinfo {author} {\bibfnamefont
  {D.}~\bibnamefont {Westly}}, \bibinfo {author} {\bibfnamefont {B.~R.}\
  \bibnamefont {Ilic}}, \bibinfo {author} {\bibfnamefont {A.}~\bibnamefont
  {Bluestone}}, \bibinfo {author} {\bibfnamefont {N.}~\bibnamefont {Volet}},
  \bibinfo {author} {\bibfnamefont {T.}~\bibnamefont {Komljenovic}}, \bibinfo
  {author} {\bibfnamefont {L.}~\bibnamefont {Chang}}, \bibinfo {author}
  {\bibfnamefont {S.~H.}\ \bibnamefont {Lee}}, \bibinfo {author} {\bibfnamefont
  {D.~Y.}\ \bibnamefont {Oh}}, \bibinfo {author} {\bibfnamefont {M.-G.}\
  \bibnamefont {Suh}}, \bibinfo {author} {\bibfnamefont {K.~Y.}\ \bibnamefont
  {Yang}}, \bibinfo {author} {\bibfnamefont {M.~H.~P.}\ \bibnamefont
  {Pfeiffer}}, \bibinfo {author} {\bibfnamefont {T.~J.}\ \bibnamefont
  {Kippenberg}}, \bibinfo {author} {\bibfnamefont {E.}~\bibnamefont {Norberg}},
  \bibinfo {author} {\bibfnamefont {L.}~\bibnamefont {Theogarajan}}, \bibinfo
  {author} {\bibfnamefont {K.}~\bibnamefont {Vahala}}, \bibinfo {author}
  {\bibfnamefont {N.~R.}\ \bibnamefont {Newbury}}, \bibinfo {author}
  {\bibfnamefont {K.}~\bibnamefont {Srinivasan}}, \bibinfo {author}
  {\bibfnamefont {J.~E.}\ \bibnamefont {Bowers}}, \bibinfo {author}
  {\bibfnamefont {S.~A.}\ \bibnamefont {Diddams}}, \ and\ \bibinfo {author}
  {\bibfnamefont {S.~B.}\ \bibnamefont {Papp}},\ }\href@noop {} {\bibfield
  {journal} {\bibinfo  {journal} {Nature}\ }\textbf {\bibinfo {volume} {557}},\
  \bibinfo {pages} {81} (\bibinfo {year} {2018})}\BibitemShut {NoStop}%
\bibitem [{\citenamefont {Dutt}\ \emph {et~al.}(2018)\citenamefont {Dutt},
  \citenamefont {Joshi}, \citenamefont {Ji}, \citenamefont {Cardenas},
  \citenamefont {Okawachi}, \citenamefont {Luke}, \citenamefont {Gaeta},\ and\
  \citenamefont {Lipson}}]{Dutt2016}%
  \BibitemOpen
  \bibfield  {author} {\bibinfo {author} {\bibfnamefont {A.}~\bibnamefont
  {Dutt}}, \bibinfo {author} {\bibfnamefont {C.}~\bibnamefont {Joshi}},
  \bibinfo {author} {\bibfnamefont {X.}~\bibnamefont {Ji}}, \bibinfo {author}
  {\bibfnamefont {J.}~\bibnamefont {Cardenas}}, \bibinfo {author}
  {\bibfnamefont {Y.}~\bibnamefont {Okawachi}}, \bibinfo {author}
  {\bibfnamefont {K.}~\bibnamefont {Luke}}, \bibinfo {author} {\bibfnamefont
  {A.~L.}\ \bibnamefont {Gaeta}}, \ and\ \bibinfo {author} {\bibfnamefont
  {M.}~\bibnamefont {Lipson}},\ }\href@noop {} {\bibfield  {journal} {\bibinfo
  {journal} {Science Advances}\ }\textbf {\bibinfo {volume} {4}} (\bibinfo
  {year} {2018})}\BibitemShut {NoStop}%
\bibitem [{\citenamefont {Suh}\ \emph {et~al.}(2016)\citenamefont {Suh},
  \citenamefont {Yang}, \citenamefont {Yang}, \citenamefont {Yi},\ and\
  \citenamefont {Vahala}}]{Suh2016}%
  \BibitemOpen
  \bibfield  {author} {\bibinfo {author} {\bibfnamefont {M.-G.}\ \bibnamefont
  {Suh}}, \bibinfo {author} {\bibfnamefont {Q.-F.}\ \bibnamefont {Yang}},
  \bibinfo {author} {\bibfnamefont {K.~Y.}\ \bibnamefont {Yang}}, \bibinfo
  {author} {\bibfnamefont {X.}~\bibnamefont {Yi}}, \ and\ \bibinfo {author}
  {\bibfnamefont {K.~J.}\ \bibnamefont {Vahala}},\ }\href@noop {} {\bibfield
  {journal} {\bibinfo  {journal} {Science}\ }\textbf {\bibinfo {volume}
  {354}},\ \bibinfo {pages} {600} (\bibinfo {year} {2016})}\BibitemShut
  {NoStop}%
\bibitem [{\citenamefont {Marin-Palomo}\ \emph {et~al.}(2017)\citenamefont
  {Marin-Palomo}, \citenamefont {Kemal}, \citenamefont {Karpov}, \citenamefont
  {Kordts}, \citenamefont {Pfeifle}, \citenamefont {Pfeiffer}, \citenamefont
  {Trocha}, \citenamefont {Wolf}, \citenamefont {Brasch}, \citenamefont
  {Anderson}, \citenamefont {Rosenberger}, \citenamefont {Vijayan},
  \citenamefont {Freude}, \citenamefont {Kippenberg},\ and\ \citenamefont
  {Koos}}]{Marin-Palomo2017}%
  \BibitemOpen
  \bibfield  {author} {\bibinfo {author} {\bibfnamefont {P.}~\bibnamefont
  {Marin-Palomo}}, \bibinfo {author} {\bibfnamefont {J.~N.}\ \bibnamefont
  {Kemal}}, \bibinfo {author} {\bibfnamefont {M.}~\bibnamefont {Karpov}},
  \bibinfo {author} {\bibfnamefont {A.}~\bibnamefont {Kordts}}, \bibinfo
  {author} {\bibfnamefont {J.}~\bibnamefont {Pfeifle}}, \bibinfo {author}
  {\bibfnamefont {M.~H.}\ \bibnamefont {Pfeiffer}}, \bibinfo {author}
  {\bibfnamefont {P.}~\bibnamefont {Trocha}}, \bibinfo {author} {\bibfnamefont
  {S.}~\bibnamefont {Wolf}}, \bibinfo {author} {\bibfnamefont {V.}~\bibnamefont
  {Brasch}}, \bibinfo {author} {\bibfnamefont {M.~H.}\ \bibnamefont
  {Anderson}}, \bibinfo {author} {\bibfnamefont {R.}~\bibnamefont
  {Rosenberger}}, \bibinfo {author} {\bibfnamefont {K.}~\bibnamefont
  {Vijayan}}, \bibinfo {author} {\bibfnamefont {W.}~\bibnamefont {Freude}},
  \bibinfo {author} {\bibfnamefont {T.~J.}\ \bibnamefont {Kippenberg}}, \ and\
  \bibinfo {author} {\bibfnamefont {C.}~\bibnamefont {Koos}},\ }\href {\doibase
  10.1038/nature22387} {\bibfield  {journal} {\bibinfo  {journal} {Nature}\
  }\textbf {\bibinfo {volume} {546}},\ \bibinfo {pages} {274} (\bibinfo {year}
  {2017})}\BibitemShut {NoStop}%
\bibitem [{\citenamefont {Trocha}\ \emph {et~al.}(2018)\citenamefont {Trocha},
  \citenamefont {Karpov}, \citenamefont {Ganin}, \citenamefont {Pfeiffer},
  \citenamefont {Kordts}, \citenamefont {Wolf}, \citenamefont {Krockenberger},
  \citenamefont {Marin-Palomo}, \citenamefont {Weimann}, \citenamefont
  {Randel}, \citenamefont {Freude}, \citenamefont {Kippenberg},\ and\
  \citenamefont {Koos}}]{Trocha2017}%
  \BibitemOpen
  \bibfield  {author} {\bibinfo {author} {\bibfnamefont {P.}~\bibnamefont
  {Trocha}}, \bibinfo {author} {\bibfnamefont {M.}~\bibnamefont {Karpov}},
  \bibinfo {author} {\bibfnamefont {D.}~\bibnamefont {Ganin}}, \bibinfo
  {author} {\bibfnamefont {M.~H.~P.}\ \bibnamefont {Pfeiffer}}, \bibinfo
  {author} {\bibfnamefont {A.}~\bibnamefont {Kordts}}, \bibinfo {author}
  {\bibfnamefont {S.}~\bibnamefont {Wolf}}, \bibinfo {author} {\bibfnamefont
  {J.}~\bibnamefont {Krockenberger}}, \bibinfo {author} {\bibfnamefont
  {P.}~\bibnamefont {Marin-Palomo}}, \bibinfo {author} {\bibfnamefont
  {C.}~\bibnamefont {Weimann}}, \bibinfo {author} {\bibfnamefont
  {S.}~\bibnamefont {Randel}}, \bibinfo {author} {\bibfnamefont
  {W.}~\bibnamefont {Freude}}, \bibinfo {author} {\bibfnamefont {T.~J.}\
  \bibnamefont {Kippenberg}}, \ and\ \bibinfo {author} {\bibfnamefont
  {C.}~\bibnamefont {Koos}},\ }\href@noop {} {\bibfield  {journal} {\bibinfo
  {journal} {Science}\ }\textbf {\bibinfo {volume} {359}},\ \bibinfo {pages}
  {887} (\bibinfo {year} {2018})}\BibitemShut {NoStop}%
\bibitem [{\citenamefont {Jost}\ \emph {et~al.}(2014)\citenamefont {Jost},
  \citenamefont {Herr}, \citenamefont {Lecaplain}, \citenamefont {Brasch},
  \citenamefont {Pfeiffer},\ and\ \citenamefont {Kippenberg}}]{Jost2014}%
  \BibitemOpen
  \bibfield  {author} {\bibinfo {author} {\bibfnamefont {J.~D.}\ \bibnamefont
  {Jost}}, \bibinfo {author} {\bibfnamefont {T.}~\bibnamefont {Herr}}, \bibinfo
  {author} {\bibfnamefont {C.}~\bibnamefont {Lecaplain}}, \bibinfo {author}
  {\bibfnamefont {V.}~\bibnamefont {Brasch}}, \bibinfo {author} {\bibfnamefont
  {M.~H.~P.}\ \bibnamefont {Pfeiffer}}, \ and\ \bibinfo {author} {\bibfnamefont
  {T.~J.}\ \bibnamefont {Kippenberg}},\ }\href {http://arxiv.org/abs/1411.1354
  http://dx.doi.org/10.1364/OPTICA.2.000706} {\bibfield  {journal} {\bibinfo
  {journal} {Optica}\ }\textbf {\bibinfo {volume} {2}} (\bibinfo {year}
  {2014})}\BibitemShut {NoStop}%
\bibitem [{\citenamefont {Brasch}\ \emph {et~al.}(2017)\citenamefont {Brasch},
  \citenamefont {Lucas}, \citenamefont {Jost}, \citenamefont {Geiselmann},\
  and\ \citenamefont {Kippenberg}}]{Brasch2017}%
  \BibitemOpen
  \bibfield  {author} {\bibinfo {author} {\bibfnamefont {V.}~\bibnamefont
  {Brasch}}, \bibinfo {author} {\bibfnamefont {E.}~\bibnamefont {Lucas}},
  \bibinfo {author} {\bibfnamefont {J.~D.}\ \bibnamefont {Jost}}, \bibinfo
  {author} {\bibfnamefont {M.}~\bibnamefont {Geiselmann}}, \ and\ \bibinfo
  {author} {\bibfnamefont {T.~J.}\ \bibnamefont {Kippenberg}},\ }\href
  {http://dx.doi.org/10.1038/lsa.2016.202} {\bibfield  {journal} {\bibinfo
  {journal} {Light: Science and Applications}\ }\textbf {\bibinfo {volume} {6}}
  (\bibinfo {year} {2017})}\BibitemShut {NoStop}%
\bibitem [{\citenamefont {Pfeiffer}\ \emph {et~al.}(2017)\citenamefont
  {Pfeiffer}, \citenamefont {Herkommer}, \citenamefont {Liu}, \citenamefont
  {Guo}, \citenamefont {Karpov}, \citenamefont {Lucas}, \citenamefont
  {Zervas},\ and\ \citenamefont {Kippenberg}}]{Pfeiffer2017}%
  \BibitemOpen
  \bibfield  {author} {\bibinfo {author} {\bibfnamefont {M.~H.~P.}\
  \bibnamefont {Pfeiffer}}, \bibinfo {author} {\bibfnamefont {C.}~\bibnamefont
  {Herkommer}}, \bibinfo {author} {\bibfnamefont {J.}~\bibnamefont {Liu}},
  \bibinfo {author} {\bibfnamefont {H.}~\bibnamefont {Guo}}, \bibinfo {author}
  {\bibfnamefont {M.}~\bibnamefont {Karpov}}, \bibinfo {author} {\bibfnamefont
  {E.}~\bibnamefont {Lucas}}, \bibinfo {author} {\bibfnamefont
  {M.}~\bibnamefont {Zervas}}, \ and\ \bibinfo {author} {\bibfnamefont {T.~J.}\
  \bibnamefont {Kippenberg}},\ }\href {http://arxiv.org/abs/1701.08594}
  {\bibfield  {journal} {\bibinfo  {journal} {Optica}\ }\textbf {\bibinfo
  {volume} {4}} (\bibinfo {year} {2017})}\BibitemShut {NoStop}%
\bibitem [{\citenamefont {Endo}\ \emph {et~al.}(2018)\citenamefont {Endo},
  \citenamefont {Shoji},\ and\ \citenamefont {Schibli}}]{Endo2018}%
  \BibitemOpen
  \bibfield  {author} {\bibinfo {author} {\bibfnamefont {M.}~\bibnamefont
  {Endo}}, \bibinfo {author} {\bibfnamefont {T.~D.}\ \bibnamefont {Shoji}}, \
  and\ \bibinfo {author} {\bibfnamefont {T.~R.}\ \bibnamefont {Schibli}},\
  }\href@noop {} {\bibfield  {journal} {\bibinfo  {journal} {IEEE Journal of
  Selected Topics in Quantum Electronics}\ }\textbf {\bibinfo {volume} {24}}
  (\bibinfo {year} {2018})}\BibitemShut {NoStop}%
\bibitem [{\citenamefont {Carlson}\ \emph {et~al.}(2018)\citenamefont
  {Carlson}, \citenamefont {Hickstein}, \citenamefont {Zhang}, \citenamefont
  {Metcalf}, \citenamefont {Quinlan}, \citenamefont {Diddams},\ and\
  \citenamefont {Papp}}]{Carlson2017}%
  \BibitemOpen
  \bibfield  {author} {\bibinfo {author} {\bibfnamefont {D.~R.}\ \bibnamefont
  {Carlson}}, \bibinfo {author} {\bibfnamefont {D.~D.}\ \bibnamefont
  {Hickstein}}, \bibinfo {author} {\bibfnamefont {W.}~\bibnamefont {Zhang}},
  \bibinfo {author} {\bibfnamefont {A.~J.}\ \bibnamefont {Metcalf}}, \bibinfo
  {author} {\bibfnamefont {F.}~\bibnamefont {Quinlan}}, \bibinfo {author}
  {\bibfnamefont {S.~A.}\ \bibnamefont {Diddams}}, \ and\ \bibinfo {author}
  {\bibfnamefont {S.~B.}\ \bibnamefont {Papp}},\ }\href@noop {} {\bibfield
  {journal} {\bibinfo  {journal} {Science}\ }\textbf {\bibinfo {volume}
  {361}},\ \bibinfo {pages} {1358} (\bibinfo {year} {2018})}\BibitemShut
  {NoStop}%
\bibitem [{\citenamefont {Stone}\ \emph {et~al.}(2018)\citenamefont {Stone},
  \citenamefont {Briles}, \citenamefont {Drake}, \citenamefont {Spencer},
  \citenamefont {Carlson}, \citenamefont {Diddams},\ and\ \citenamefont
  {Papp}}]{Stone2017}%
  \BibitemOpen
  \bibfield  {author} {\bibinfo {author} {\bibfnamefont {J.~R.}\ \bibnamefont
  {Stone}}, \bibinfo {author} {\bibfnamefont {T.~C.}\ \bibnamefont {Briles}},
  \bibinfo {author} {\bibfnamefont {T.~E.}\ \bibnamefont {Drake}}, \bibinfo
  {author} {\bibfnamefont {D.~T.}\ \bibnamefont {Spencer}}, \bibinfo {author}
  {\bibfnamefont {D.~R.}\ \bibnamefont {Carlson}}, \bibinfo {author}
  {\bibfnamefont {S.~A.}\ \bibnamefont {Diddams}}, \ and\ \bibinfo {author}
  {\bibfnamefont {S.~B.}\ \bibnamefont {Papp}},\ }\href {\doibase
  10.1103/PhysRevLett.121.063902} {\bibfield  {journal} {\bibinfo  {journal}
  {Phys. Rev. Lett.}\ }\textbf {\bibinfo {volume} {121}},\ \bibinfo {pages}
  {063902} (\bibinfo {year} {2018})}\BibitemShut {NoStop}%
\bibitem [{\citenamefont {Del'Haye}\ \emph {et~al.}(2012)\citenamefont
  {Del'Haye}, \citenamefont {Papp},\ and\ \citenamefont
  {Diddams}}]{delhaye2012}%
  \BibitemOpen
  \bibfield  {author} {\bibinfo {author} {\bibfnamefont {P.}~\bibnamefont
  {Del'Haye}}, \bibinfo {author} {\bibfnamefont {S.~B.}\ \bibnamefont {Papp}},
  \ and\ \bibinfo {author} {\bibfnamefont {S.~A.}\ \bibnamefont {Diddams}},\
  }\href@noop {} {\bibfield  {journal} {\bibinfo  {journal} {Phys. Rev. Lett.}\
  }\textbf {\bibinfo {volume} {109}},\ \bibinfo {pages} {263901} (\bibinfo
  {year} {2012})}\BibitemShut {NoStop}%
\bibitem [{\citenamefont {Wilson}\ \emph {et~al.}(2015)\citenamefont {Wilson},
  \citenamefont {Nadeau}, \citenamefont {Jaworski}, \citenamefont {Tromberg},\
  and\ \citenamefont {Durkin}}]{Wilson2015}%
  \BibitemOpen
  \bibfield  {author} {\bibinfo {author} {\bibfnamefont {R.~H.}\ \bibnamefont
  {Wilson}}, \bibinfo {author} {\bibfnamefont {K.~P.}\ \bibnamefont {Nadeau}},
  \bibinfo {author} {\bibfnamefont {F.~B.}\ \bibnamefont {Jaworski}}, \bibinfo
  {author} {\bibfnamefont {B.~J.}\ \bibnamefont {Tromberg}}, \ and\ \bibinfo
  {author} {\bibfnamefont {A.~J.}\ \bibnamefont {Durkin}},\ }\href {\doibase
  10.1117/1.JBO.20.3.030901} {\bibfield  {journal} {\bibinfo  {journal}
  {Journal of Biomedical Optics}\ }\textbf {\bibinfo {volume} {20}},\ \bibinfo
  {pages} {030901} (\bibinfo {year} {2015})}\BibitemShut {NoStop}%
\bibitem [{\citenamefont {Manley}(2014)}]{Manley2014}%
  \BibitemOpen
  \bibfield  {author} {\bibinfo {author} {\bibfnamefont {M.}~\bibnamefont
  {Manley}},\ }\href {\doibase 10.1039/C4CS00062E} {\bibfield  {journal}
  {\bibinfo  {journal} {Chem. Soc. Rev.}\ }\textbf {\bibinfo {volume} {43}},\
  \bibinfo {pages} {8200} (\bibinfo {year} {2014})}\BibitemShut {NoStop}%
\bibitem [{\citenamefont {Burgess}\ and\ \citenamefont
  {Hammond}(2007)}]{Burgess2007}%
  \BibitemOpen
  \bibfield  {author} {\bibinfo {author} {\bibfnamefont {C.}~\bibnamefont
  {Burgess}}\ and\ \bibinfo {author} {\bibfnamefont {J.}~\bibnamefont
  {Hammond}},\ }\href {www.spectroscopyonline.com} {\bibfield  {journal}
  {\bibinfo  {journal} {Spectroscopy}\ }\textbf {\bibinfo {volume} {22}},\
  \bibinfo {pages} {40} (\bibinfo {year} {2007})}\BibitemShut {NoStop}%
\bibitem [{\citenamefont {Rohe}\ \emph {et~al.}(1999)\citenamefont {Rohe},
  \citenamefont {Becker}, \citenamefont {K{\"{o}}lle}, \citenamefont
  {Eisenreich},\ and\ \citenamefont {Eyerer}}]{Rohe1999}%
  \BibitemOpen
  \bibfield  {author} {\bibinfo {author} {\bibfnamefont {T.}~\bibnamefont
  {Rohe}}, \bibinfo {author} {\bibfnamefont {W.}~\bibnamefont {Becker}},
  \bibinfo {author} {\bibfnamefont {S.}~\bibnamefont {K{\"{o}}lle}}, \bibinfo
  {author} {\bibfnamefont {N.}~\bibnamefont {Eisenreich}}, \ and\ \bibinfo
  {author} {\bibfnamefont {P.}~\bibnamefont {Eyerer}},\ }\href {\doibase
  10.1016/S0039-9140(99)00035-1} {\bibfield  {journal} {\bibinfo  {journal}
  {Talanta}\ }\textbf {\bibinfo {volume} {50}},\ \bibinfo {pages} {283}
  (\bibinfo {year} {1999})}\BibitemShut {NoStop}%
\bibitem [{\citenamefont {Tsuchikawa}\ \emph {et~al.}(2005)\citenamefont
  {Tsuchikawa}, \citenamefont {Yonenobu},\ and\ \citenamefont
  {Siesler}}]{Tsuchikawa2005}%
  \BibitemOpen
  \bibfield  {author} {\bibinfo {author} {\bibfnamefont {S.}~\bibnamefont
  {Tsuchikawa}}, \bibinfo {author} {\bibfnamefont {H.}~\bibnamefont
  {Yonenobu}}, \ and\ \bibinfo {author} {\bibfnamefont {H.~W.}\ \bibnamefont
  {Siesler}},\ }\href {\doibase 10.1039/b412759e} {\bibfield  {journal}
  {\bibinfo  {journal} {The Analyst}\ }\textbf {\bibinfo {volume} {130}},\
  \bibinfo {pages} {379} (\bibinfo {year} {2005})}\BibitemShut {NoStop}%
\bibitem [{\citenamefont {Jensen}\ and\ \citenamefont
  {Glotch}(2011)}]{Jensen2011}%
  \BibitemOpen
  \bibfield  {author} {\bibinfo {author} {\bibfnamefont {H.~B.}\ \bibnamefont
  {Jensen}}\ and\ \bibinfo {author} {\bibfnamefont {T.~D.}\ \bibnamefont
  {Glotch}},\ }\href@noop {} {\bibfield  {journal} {\bibinfo  {journal}
  {Journal of Geophysical Research E: Planets}\ }\textbf {\bibinfo {volume}
  {116}} (\bibinfo {year} {2011})}\BibitemShut {NoStop}%
\bibitem [{\citenamefont {Delaney}\ \emph {et~al.}(2005)\citenamefont
  {Delaney}, \citenamefont {Walmsley}, \citenamefont {Berrie},\ and\
  \citenamefont {Fletcher}}]{Delaney2005}%
  \BibitemOpen
  \bibfield  {author} {\bibinfo {author} {\bibfnamefont {J.~K.~J.}\
  \bibnamefont {Delaney}}, \bibinfo {author} {\bibfnamefont {E.}~\bibnamefont
  {Walmsley}}, \bibinfo {author} {\bibfnamefont {B.~H.}\ \bibnamefont
  {Berrie}}, \ and\ \bibinfo {author} {\bibfnamefont {C.~F.}\ \bibnamefont
  {Fletcher}},\ }in\ \href {\doibase 10.17226/11413} {\emph {\bibinfo
  {booktitle} {Scientific Examination of Art: Modern Techniques in Conservation
  and Analysis - Proceedings of the National Academy of Sciences National
  Academies Press, Aug 16, 2005, (Sackler NAS Colloquium)}}}\ (\bibinfo {year}
  {2005})\ pp.\ \bibinfo {pages} {120--136}\BibitemShut {NoStop}%
\bibitem [{\citenamefont {Dooley}\ \emph {et~al.}(2013)\citenamefont {Dooley},
  \citenamefont {Lomax}, \citenamefont {Zeibel}, \citenamefont {Miliani},
  \citenamefont {Ricciardi}, \citenamefont {Hoenigswald}, \citenamefont
  {Loew},\ and\ \citenamefont {Delaney}}]{Dooley2013}%
  \BibitemOpen
  \bibfield  {author} {\bibinfo {author} {\bibfnamefont {K.~A.}\ \bibnamefont
  {Dooley}}, \bibinfo {author} {\bibfnamefont {S.}~\bibnamefont {Lomax}},
  \bibinfo {author} {\bibfnamefont {J.~G.}\ \bibnamefont {Zeibel}}, \bibinfo
  {author} {\bibfnamefont {C.}~\bibnamefont {Miliani}}, \bibinfo {author}
  {\bibfnamefont {P.}~\bibnamefont {Ricciardi}}, \bibinfo {author}
  {\bibfnamefont {A.}~\bibnamefont {Hoenigswald}}, \bibinfo {author}
  {\bibfnamefont {M.}~\bibnamefont {Loew}}, \ and\ \bibinfo {author}
  {\bibfnamefont {J.~K.}\ \bibnamefont {Delaney}},\ }\href {\doibase
  10.1039/c3an00926b} {\bibfield  {journal} {\bibinfo  {journal} {The Analyst}\
  }\textbf {\bibinfo {volume} {138}},\ \bibinfo {pages} {4838} (\bibinfo {year}
  {2013})}\BibitemShut {NoStop}%
\bibitem [{\citenamefont {Palmer}\ and\ \citenamefont
  {Williams}(1974)}]{Palmer1974}%
  \BibitemOpen
  \bibfield  {author} {\bibinfo {author} {\bibfnamefont {K.~F.}\ \bibnamefont
  {Palmer}}\ and\ \bibinfo {author} {\bibfnamefont {D.}~\bibnamefont
  {Williams}},\ }\href {\doibase 10.1364/JOSA.64.001107} {\bibfield  {journal}
  {\bibinfo  {journal} {Journal of the Optical Society of America}\ }\textbf
  {\bibinfo {volume} {64}},\ \bibinfo {pages} {1107} (\bibinfo {year}
  {1974})}\BibitemShut {NoStop}%
\bibitem [{\citenamefont {van Veen}\ \emph {et~al.}(2004)\citenamefont {van
  Veen}, \citenamefont {Sterenborg}, \citenamefont {Pifferi}, \citenamefont
  {Torricelli},\ and\ \citenamefont {Cubeddu}}]{Veen2004}%
  \BibitemOpen
  \bibfield  {author} {\bibinfo {author} {\bibfnamefont {R.~L.}\ \bibnamefont
  {van Veen}}, \bibinfo {author} {\bibfnamefont {H.}~\bibnamefont
  {Sterenborg}}, \bibinfo {author} {\bibfnamefont {A.}~\bibnamefont {Pifferi}},
  \bibinfo {author} {\bibfnamefont {A.}~\bibnamefont {Torricelli}}, \ and\
  \bibinfo {author} {\bibfnamefont {R.}~\bibnamefont {Cubeddu}},\ }\href@noop
  {} {\bibfield  {journal} {\bibinfo  {journal} {Biomedical Topical Meeting,
  OSA Technical Digest}\ } (\bibinfo {year} {2004})}\BibitemShut {NoStop}%
\bibitem [{\citenamefont {Stanton}\ \emph {et~al.}(2015)\citenamefont
  {Stanton}, \citenamefont {Heck}, \citenamefont {Bovington}, \citenamefont
  {Spott},\ and\ \citenamefont {Bowers}}]{Stanton2015}%
  \BibitemOpen
  \bibfield  {author} {\bibinfo {author} {\bibfnamefont {E.~J.}\ \bibnamefont
  {Stanton}}, \bibinfo {author} {\bibfnamefont {M.~J.~R.}\ \bibnamefont
  {Heck}}, \bibinfo {author} {\bibfnamefont {J.}~\bibnamefont {Bovington}},
  \bibinfo {author} {\bibfnamefont {A.}~\bibnamefont {Spott}}, \ and\ \bibinfo
  {author} {\bibfnamefont {J.~E.}\ \bibnamefont {Bowers}},\ }\href {\doibase
  10.1364/OE.23.011272} {\bibfield  {journal} {\bibinfo  {journal} {Optics
  Express}\ }\textbf {\bibinfo {volume} {23}},\ \bibinfo {pages} {11272}
  (\bibinfo {year} {2015})}\BibitemShut {NoStop}%
\bibitem [{\citenamefont {Gatkine}\ \emph {et~al.}(2017)\citenamefont
  {Gatkine}, \citenamefont {Veilleux}, \citenamefont {Hu}, \citenamefont
  {Bland-Hawthorn},\ and\ \citenamefont {Dagenais}}]{Gatkine2017}%
  \BibitemOpen
  \bibfield  {author} {\bibinfo {author} {\bibfnamefont {P.}~\bibnamefont
  {Gatkine}}, \bibinfo {author} {\bibfnamefont {S.}~\bibnamefont {Veilleux}},
  \bibinfo {author} {\bibfnamefont {Y.}~\bibnamefont {Hu}}, \bibinfo {author}
  {\bibfnamefont {J.}~\bibnamefont {Bland-Hawthorn}}, \ and\ \bibinfo {author}
  {\bibfnamefont {M.}~\bibnamefont {Dagenais}},\ }\href {\doibase
  10.1364/OE.25.017918} {\bibfield  {journal} {\bibinfo  {journal} {Optics
  Express}\ }\textbf {\bibinfo {volume} {25}},\ \bibinfo {pages} {232}
  (\bibinfo {year} {2017})}\BibitemShut {NoStop}%
\bibitem [{\citenamefont {Stanton}\ \emph {et~al.}(2017)\citenamefont
  {Stanton}, \citenamefont {Volet},\ and\ \citenamefont
  {Bowers}}]{Stanton2017}%
  \BibitemOpen
  \bibfield  {author} {\bibinfo {author} {\bibfnamefont {E.~J.}\ \bibnamefont
  {Stanton}}, \bibinfo {author} {\bibfnamefont {N.}~\bibnamefont {Volet}}, \
  and\ \bibinfo {author} {\bibfnamefont {J.~E.}\ \bibnamefont {Bowers}},\
  }\href {\doibase 10.1364/OE.25.030651} {\bibfield  {journal} {\bibinfo
  {journal} {Optics Express}\ }\textbf {\bibinfo {volume} {25}},\ \bibinfo
  {pages} {30651} (\bibinfo {year} {2017})}\BibitemShut {NoStop}%
\end{thebibliography}%

\end{document}